\documentclass[journal=jctcce,manuscript=article]{achemso}

\usepackage[version=3]{mhchem} 
\usepackage{xcolor}
\usepackage{amsfonts} 
\usepackage{caption}
\usepackage{subcaption}
\usepackage[capitalise]{cleveref}

\newcommand{\vk}{\mathbf{k}}
\newcommand{\vq}{\mathbf{q}}
\newcommand{\vG}{\mathbf{G}}

\setkeys{acs}{maxauthors = 3}

\author{Stephen Jon Quiton}
\author{Hamlin Wu}
\affiliation{College of Chemistry, University of California, Berkeley, California 94720, United States}

\author{Xin Xing}
\author{Lin Lin}
\email{linlin@math.berkeley.edu}
\affiliation{Department of Mathematics, University of California, Berkeley, California 94720, United States}

\author{Martin Head-Gordon}
\email{mhg@cchem.berkeley.edu}
\affiliation{College of Chemistry, University of California, Berkeley, California 94720, United States}

\title[]{The Staggered Mesh Method: Accurate Exact Exchange towards the Thermodynamic Limit for Solids}


\begin{document}

\begin{tocentry}

    
\includegraphics[width=\textwidth]{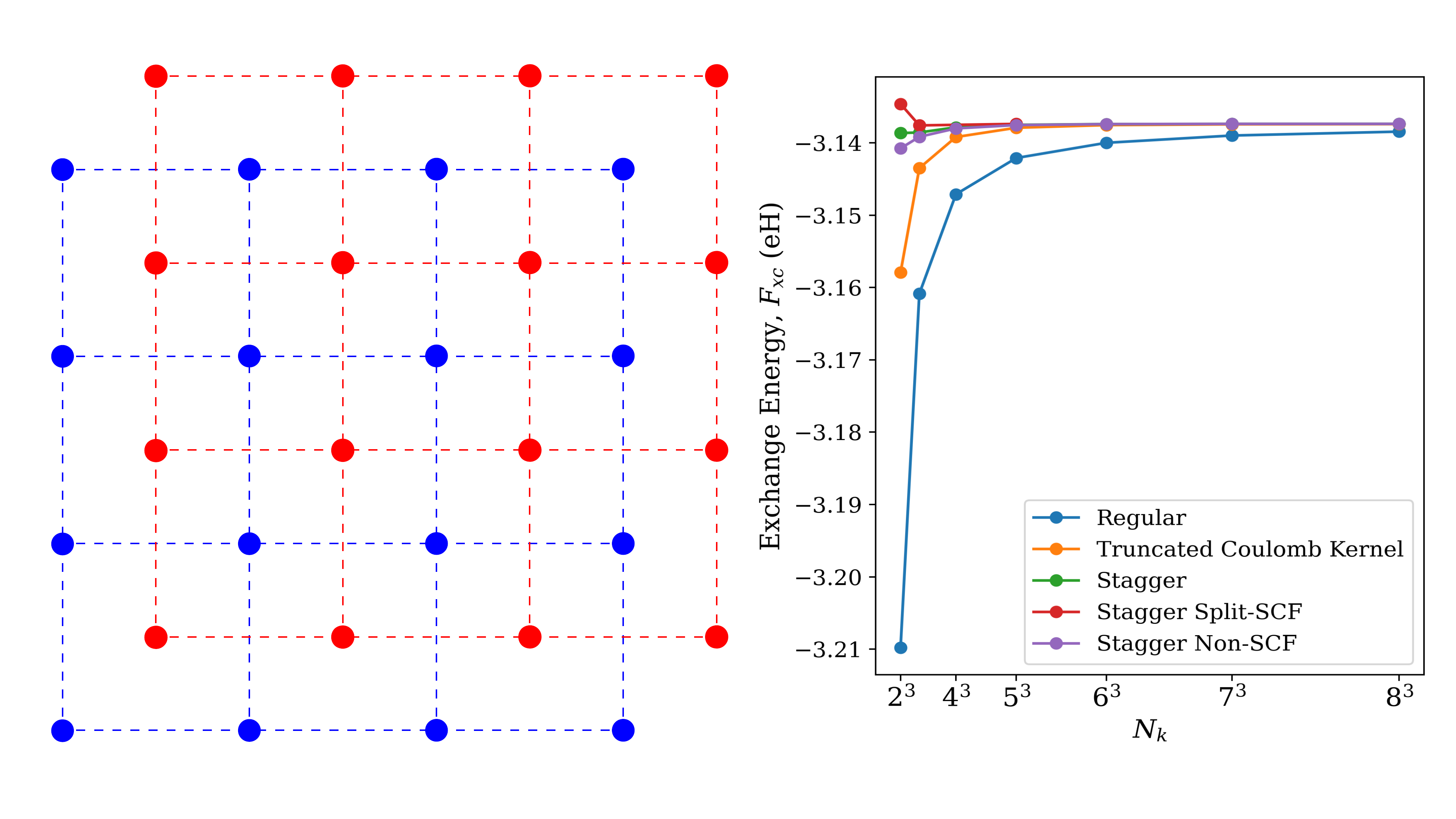}




\end{tocentry}

\begin{abstract}
In periodic systems, the Hartree-Fock (HF) exchange energy exhibits the slowest convergence of all HF energy components as the system size approaches the thermodynamic limit. We demonstrate that the recently proposed staggered mesh method for Fock exchange energy [Xing, Li, and Lin, Math. Comp., 2024], which is specifically designed to sidestep certain singularities in exchange energy evaluation, can expedite the finite-size convergence rate for the exact exchange energy across a range of insulators and semiconductors when compared to the regular and truncated Coulomb methods. This remains true even for two computationally cheaper versions of this new method, which we call Non-SCF and Split-SCF staggered mesh. Additionally, a sequence of numerical tests on simple solids showcases the staggered mesh method's ability to improve convergence towards the thermodynamic limit for band gaps, bulk moduli, equilibrium lattice dimensions, energies, and phonon force constants.

\end{abstract}

\section{Introduction}


Wavefunction-based methods are becoming increasingly popular for use in electronic structure calculations of solid-state materials.\cite{booth_towards_2013,raabe_dont_2002,pisani_periodic_2008,huang_advances_2008,jain_commentary_2013}. However, a longstanding, inherent problem with these methods is the finite-size error; specifically, what \textit{finite} size of the periodic lattice is required to reach chemical accuracy with respect to the simulation of the \textit{infinite} lattice, that is, the thermodynamic limit (TDL)? 
One approach to this problem is to simply increase the number of unit cells along the lattice dimensions; 
however, doing so will increase the number of electrons, atoms, and basis functions,  in the unit cell resulting in greater computational expense. When performing periodic Hartree-Fock (HF) calculations, which is used as a reference for wavefunction methods and used in exact exchange for hybrid Density Functional Theory (DFT), many solid-state quantum chemistry codes use an equivalent yet more efficient approach called \textbf{k}-point sampling, where \textbf{k} is a momentum vector in a Bloch wavefunction. 
In this approach, the TDL is defined as the limit when the number of \textbf{k}-points sampled over the First Brillouin Zone (FBZ) increases towards infinity.

The standard approach in sampling the FBZ is to use the  Monkhorst-Pack (MP) mesh \cite{monkhorst_special_1976}, which is constructed to evenly sample the FBZ given a number of points to sample along each dimension. Using this mesh, the error in the HF exchange energy as one approaches the TDL decreases as $\mathcal{O}(N_\mathrm{k}^{-1/3})$ with an increasing number of \textbf{k}-points $N_\mathrm{k}$.
 This slow convergence rate is primarily due to the excluded divergent term 
when $|\mathbf{q}+\mathbf{G}|=\mathbf{0}$ in the unscreened Coulomb kernel \cite{xing_unified_2023}. To reduce the error due to the Coulomb singularity, a common practice is to include the Madelung constant correction \cite{fraser_finite-size_1996,drummond_finite-size_2008}. Interpreting the Madelung constant correction as a singularity subtraction technique allows for a rigorous proof that the finite-size error reduces to $\mathcal{O}(N_\mathrm{k}^{-1})$ \cite{xing_unified_2023}. 

To reduce the number of \textbf{k}-points sampled and thus the computation time, one can leverage the symmetry of a unit cell for more efficient sampling. For example, sampling all \textbf{k}-points via the MP mesh in a unit cell with high space-group symmetry is redundant; only \textbf{k}-points in the irreducible Brillouin zone (IBZ) need to be sampled, and therefore the computation time can be reduced by a factor approximately equal to the number of symmetry operators in the point group.\cite{erba_crystal23_2023,orlando_full_2014,dovesi_role_1986}. The use of generalized regular \textbf{k}-point grids \cite{moreno_optimal_1992,morgan_efficiency_2018,hart_robust_2019}, or even a simple shift of the $\Gamma$-centered MP mesh can be used to accomplish this. 
However, these methods reduce the calculation runtime by a prefactor that is strongly dependent upon the level of symmetry within the unit cell. Modifications to the Coulomb kernel, including truncation\cite{spencer_efficient_2008} and averaging\cite{schafer_sampling_2023}, as well as resolution-of-identity schemes\cite{bussy_efficient_2024} have also been used to either reduce the number of \textbf{k}-points sampled increase computational efficiency when computing exact exchange towards the TDL.


The staggered mesh method is a simple alternative approach for accelerating convergence towards the TDL. Recent studies have demonstrated its effectiveness in improving the \textit{asymptotic} convergence towards TDL for HF Exchange, Second-order Moller-Plesset Perturbation Theory (MP2), and the Random Phase Approximation (RPA)
\cite{xing_unified_2023,xing_staggered_2021,xing_staggered_2022}. 
Consider the two-electron integral in reciprocal space; in the integrand of periodic HF exchange is a denominator that depends on $\left|\mathbf{q}\right|=\left|\mathbf{k}_j-\mathbf{k}_i\right|$, which is the difference between the $\mathbf{k}$-points of the two Bloch orbitals in a pair density. We exploit the fact that if the two sets of \textbf{k}-points are staggered by a half \textbf{k}-mesh size, then the term with $\mathbf{q}=\mathbf{0}$ is no longer included in the expression for the exchange energy. Under the condition of removable discontinuity,  the convergence rate of HF exchange energy to the TDL can be accelerated to $\mathcal{O}(N_\mathrm{k}^{-5/3})$ [Corollary 6.7, Ref. \citenum{xing_unified_2023}].  This acceleration has been demonstrated using a model potential.\cite{xing_unified_2023}


In this work, we compare the performance of the staggered mesh method, along with two computationally cheaper variants, with the regular method and the Truncated Coulomb method \cite{spencer_efficient_2008,guidon_robust_2009} in reducing finite size errors for computing HF exchange energies, band gaps, several insulators. We find that all three versions of the staggered mesh method are competitive
in computing HF exchange energies and band gaps, with the original  staggered mesh method showing a distinct advantage in computing phonon force constants, lattice parameters, and bulk moduli. 
Furthermore, this method of \textbf{k}-point selection is highly appealing in that it requires minimal modification to most solid-state quantum chemistry packages that are capable of band structure calculations. 

\section{Theory}
One way to approach the TDL in solid-state calculations is to use the supercell model, where one explicitly repeats the unit cell in one or more directions. For this work, however, we use \textbf{k}-point sampling of the FBZ in reciprocal space, which is equivalent to the supercell model.
We focus primarily on the convergence of the exchange energy $E_\mathrm{K}$ with respect to $N_\textbf{\text{k}}$, which contributes most to the finite size error in the HF theory. For notation, AOs are denoted by $\mu,\nu,\lambda,\sigma$, occupied MOs are denoted by $i,j,k,l$, and virtual MOs are denoted by $a,b,c,d$.

We use the Bloch orbital framework: 
\begin{equation*}
    \psi_{\mu_{\mathbf{k}}}(\mathbf{r})=\frac{1}{\sqrt{N_\textbf{k}}} \sum_{\mathbf{R}\in\mathbb{L}} \mathrm{e}^{\mathrm{i} \mathbf{k} \cdot \mathbf{R}} \phi_\mu(\mathbf{r}-\mathbf{R}),
\end{equation*}
where $\mathbf{R}$ is a vector in the real space Bravais-lattice, $\mathbb{L}$, and $\phi_\mu$ is an atomic orbital (AO). Consider a unit cell $\Omega$ with volume $\left|\Omega\right|$ and a corresponding supercell $\Omega^s$ of volume $\left|\Omega^s\right|=N_\textbf{k}\left|\Omega\right|$. Furthermore, define the pair density as 
\begin{equation*}
\rho_{\mu_{\mathbf{k}_{\mu}}\nu_{\mathbf{k}_{\nu}}}(\mathbf{r})=\phi_{\mu_{\mathbf{k}_{\mu}}}^{*}\left(\mathbf{r}\right)\phi_{\nu_{\mathbf{k}_{\nu}}}\left(\mathbf{r}\right):=\frac{1}{|\Omega|}\sum_{\mathbf{G}\in\mathbb{L}^{*}}\hat{\rho}_{\mu_{\mathbf{k}_{\mu}}\nu_{\mathbf{k}_{\nu}}}(\mathbf{G})e^{\mathrm{i}\mathbf{G}\cdot\mathbf{r}},
\end{equation*}
where $\hat{\rho}_{\mu_{\mathbf{k}_{\mu}}\nu_{\mathbf{k}_{\nu}}}(\mathbf{G})$ is the Fourier transform of the pair density. Then the two-electron integral in the AO basis is defined as 
\begin{align}
\left(\nu_{\mathbf{k}_{\nu}}\lambda_{\mathbf{k}_{\lambda}}\mid\sigma_{\mathbf{k}_{\sigma}}\mu_{\mathbf{k}_{\mu}}\right)
&=\int_{\Omega^{s}}\mathrm{~d}\mathbf{r}_{1}\int_{\Omega^{s}}\mathrm{~d}\mathbf{r}_{2}\psi_{\nu_{\mathbf{k}_{\nu}}}^{*}\left(\mathbf{r}_{1}\right)\psi_{\lambda_{\mathbf{k}_{\lambda}}}\left(\mathbf{r}_{1}\right)(\mathbf{G})V_{\text{coul}}\left(\mathbf{r}_{1},\mathbf{r}_{2}\right)\psi_{\sigma_{\mathbf{k}_{\sigma}}}^{*}\left(\mathbf{r}_{2}\right)\psi_{\mu_{\mathbf{k}_{\mu}}}\left(\mathbf{r}_{2}\right)
\nonumber\\
&=\frac{1}{\left|\Omega^{S}\right|}\sideset{}{'}\sum_{\mathbf{G}\in\mathbb{L}^{*}}\hat{\rho}_{\nu_{\mathbf{k}_{\nu}}\lambda_{\mathbf{k}_{\lambda}}}(\mathbf{G})\hat{V}_{\mathrm{coul}}\left(\mathbf{q}+\mathbf{G}\right)\hat{\rho}_{\sigma_{\mathbf{k}_{\sigma}}\mu_{\mathbf{k}_{\mu}}}\left(\mathbf{G}_{\mathbf{k}_{\lambda},\mathbf{k}_{\mu}}^{\mathbf{k}_{\nu},\mathbf{k}_{\sigma}}-\mathbf{G}\right).
\label{eq:eri-reciprocal}
\end{align}
Here, $\mathbf{G}_{\mathbf{k}_{\lambda},\mathbf{k}_{\mu}}^{\mathbf{k}_{\nu},\mathbf{k}_{\sigma}}:=\mathbf{k}_{\lambda}+\mathbf{k}_{\mu}-\mathbf{k}_{\nu}-\mathbf{k}_{\sigma}\in \mathbb{L}^*$ by crystal momentum conservation,  $\mathbf{q}=\mathbf{k}_\lambda-\mathbf{k}_\nu$ is the \textbf{k}-point difference, $\sum'$ denotes that the term where 
$\mathbf{q}+\mathbf{G}=\mathbf{0}$ is excluded, and $V_{\text{coul}}(\mathbf{r})$ and   $\hat{V}_{\text{coul}}(\mathbf{G})$ are the Coulomb kernel in real and reciprocal space respectively. We use both the default periodic Coulomb kernel in 3D 
$\hat{V}_{\text{coul}}(\mathbf{G})=\frac{4\pi}{|\mathbf{q}+\mathbf{G}|^{2}}$ 
and the truncated coulomb kernel \cite{spencer_efficient_2008,guidon_robust_2009}. 
If we define $C_{\mu i}^\mathbf{k}$ to be the MO-coefficient matrix, 
\begin{align*}
    \psi_{i_{\mathbf{k}}}(\mathbf{r})=\sum_{\mu\in n_{ao}}C_{\mu i}^{\mathbf{k}}\psi_{\mu_{\mathbf{k}}}(\mathbf{r}),
\end{align*}
then the density matrix at that same \textbf{k}, $P^\mathbf{k}$, is defined as
\begin{equation}
    P_{\mu \nu}^{\mathbf{k}}=\sum_{i \in \mathrm{occ}} C_{\mu i}^{\mathbf{k}}\left(C_{\nu i}^{\mathbf{k}}\right)^*.
\end{equation}

\subsection{Fock exchange energy and its Madelung constant correction}
With a finite-size mesh $\mathcal{K}$, the Fock exchange energy per unit cell is computed as 
\begin{equation}\label{eqn:exchange_nk}
    E_{K}^{N_\vk} =
    \dfrac{1}{N_\vk}
    \left(
    -\frac{1}{2}
    \sum_{i,j}\sum_{\vk_i, \vk_j\in\mathcal{K}}
    (i_{\vk_i}, j_{\vk_j} | j_{\vk_j}, i_{\vk_i})
    \right).
\end{equation}
In the TDL with $\mathcal{K}$ converging to $\Omega^*$ and $N_\vk \rightarrow\infty$, the energy converges as
\begin{equation}\label{eqn:exchange_tdl}
    E_{K}^{\text{TDL}}
    =
    \lim_{N_\vk\rightarrow\infty} E_K^{N_\vk}
    =
    \int_{\Omega^{*}}d\mathbf{k}_{i}\int_{\Omega^{*}}d\mathbf{k}_{j}\  \sum_{i,j} F_{K}^{ij}\left(\mathbf{k}_{j},\mathbf{k}_{i}\right),
\end{equation}
where the integrand is 
\begin{equation*}
    F_{K}^{ij}(\vk_j, \vk_i)=-\frac{N_\vk}{2|\Omega^{*}|^{2}}\sum_{i,j}
    (i_{\vk_i}, j_{\vk_j} | j_{\vk_j}, i_{\vk_i}).
\end{equation*}
Note that each of the ERIs above consists of an $N_\vk^{-1}$ factor and the overall integrand is independent of $N_\vk$. 
From a numerical quadrature perspective, the finite-size energy $E_K^{N_\vk}$ approximates the TDL energy $E_K^\text{TDL}$ by discretizing the integral of $\vk_i$ and $\vk_j$ over $\Omega^*\times \Omega^*$ 
by a $\mathcal{K}\times \mathcal{K}$ uniform mesh i.e., a trapezoidal rule. 
By studying the singularity structure of the integrand $F_X^{j,i}(\vk_j, \vk_i)$, Ref.~\citenum{xing_unified_2023} provides
a tight estimate of this finite-size error as 
\[
    \left| E_K^\text{TDL} - E_K^{N_\vk}\right| = \mathcal{O}(N_\vk^{-\frac13}).
\]

To further reduce this error, a common approach is to introduce the Madelung constant correction \cite{fraser_finite-size_1996,drummond_finite-size_2008,martin_coulomb_2004}
to the Coulomb kernel involved in the $E_K^{N_\vk}$ calculation as 
\begin{align}
    \hat{V}_{\text {coul}}(\mathbf{G})= 
    \begin{cases}
    \frac{4 \pi}{|\mathbf{G}|^2} & \mathbf{G} \neq \mathbf{0} \\ 
    -|\Omega| N_{\mathbf{k}} \xi & \mathbf{G}=\mathbf{0}
    \end{cases},
\end{align}
which leads to a Madelung-corrected exchange energy calculation as 
\begin{equation}\label{eqn:exchange_madelung}
    E_K^{N_\textbf{k},\xi} = E_K^{N_\textbf{k}}+n_{\mathrm{occ}}\xi.
\end{equation}
The Madelung constant $\xi$ above is defined as 
\begin{align}
\xi
&=
\left(\frac{|\Omega^*|}{N_{\mathbf{k}}}\sum_{\mathbf{q}\in\mathcal{K}_{\mathbf{q}}}-\int_{\Omega^{*}}d\mathbf{q}\right)
\left(\frac{1}{(2\pi)^{3}}\sideset{}{'}\sum_{\mathbf{G}\in\mathbb{L}^{*}}\frac{4\pi e^{-|\mathbf{q}+\mathbf{G}|^{2}/\eta}}{|\mathbf{q}+\mathbf{G}|^{2}}\right)
+\sideset{}{'}\sum_{\mathbf{R}\in\mathbb{L}_{\mathcal{K}_{\mathbf{q}}}}\frac{\operatorname{erfc}\left(\eta^{1/2}|\mathbf{R}|/2\right)}{|\mathbf{R}|}-\frac{4\pi}{|\Omega|N_{\mathbf{k}}\eta},
\label{eq:madelung-regular}
\end{align}
where $\mathcal{K}_\mathbf{q}$ is a uniform mesh in $\Omega^*$ that contains the $\Gamma$-point and has the same size as $\mathcal{K}$, and $\mathbb{L}_{\mathcal{K}_\mathbf{q}}$ is the real space lattice associated with all 
points $\mathbf{q}+\mathbf{G}$ in reciprocal space with $\mathbf{q}\in\mathcal{K}_\mathbf{q}$, $\mathbf{G}\in \mathbb{L}^*$. 
Note that $\xi$ is a constant for any arbitrary $\eta>0$. 
Connecting to the equivalent supercell model, $\mathbb{L}_{\mathcal{K}_\mathbf{q}}$ is exactly the Bravais lattice of the supercell and $\vq+\vG$ defines the associated reciprocal lattice. 
This method may also be physically interpreted as placing a probe electron at the origin in a uniform charge-compensating background and calculating the potential energy due to interactions with the electron's periodic images via Ewald summation\cite{hub_quantifying_2014,paier_perdewburkeernzerhof_2005}.

The Madelung constant correction reduces the finite-size error of the exchange energy to $\mathcal{O}(N_\vk^{-1})$\cite{xing_unified_2023}.
The effectiveness of this correction can be rigorously justified from a numerical quadrature perspective. 
It turns out that $\vq$ in \eqref{eq:madelung-regular} can be interpreted as the momentum transfer vector $\vk_i - \vk_j$ in the exchange energy calculation \eqref{eqn:exchange_nk}, which also explains the definition of $\mathcal{K}_\vq$ in $\xi$. 
Plugging the definition of $\xi$ into \eqref{eqn:exchange_madelung}, the two terms in the bracket of \eqref{eq:madelung-regular} effectively replace the numerical quadrature of the leading singular term in the 
integrand $F^{ij}(\vk_i, \vk_j)$ by its exact integral. 
As a result, the Madelung constant correction improves the integrand smoothness in the finite-size exchange energy calculation (viewed as a numerical quadrature) and thus results in quadrature error reduction. 
Such a correction is connected to a common numerical quadrature technique for singular integrals called the singularity subtraction method.

\subsection{Staggered Mesh Method for Fock exchange}
To approximate the TDL exchange energy as an integral in \eqref{eqn:exchange_tdl}, a general finite-size calculation is to discretize the integrals of $\vk_i$ and $\vk_j$ separately using two 
arbitrary uniform meshes $\mathcal{K}_i$ and $\mathcal{K}_j$, i.e., 
\begin{align}
E_{K}^{N_{\textbf{k}}}\left(\mathcal{K}_{i},\mathcal{K}_{j}\right)
&=\dfrac{|\Omega^*|^2}{N_\vk^2}\sum_{\mathbf{k}_{i}\in\mathcal{K}_{i},\mathbf{k}_{j}\in\mathcal{K}_{j}}\sum_{i,j}F^{ij}_K(\vk_i, \vk_j)
\nonumber\\
&=\frac{1}{N_\vk}
\left(
-\frac12
\sum_{\mathbf{k}_{i}\in\mathcal{K}_{i},\mathbf{k}_{j}\in\mathcal{K}_{j}}\sum_{i,j}\left(j_{\mathbf{k}_{j}}i_{\mathbf{k}_{i}}\mid i_{\mathbf{k}_{i}}j_{\mathbf{k}_{j}}\right)
\right).
\label{eq:ek_mo}
\end{align}
In this setting, the regular exchange energy calculation $E_K^{N_\vk}$ in \eqref{eqn:exchange_nk} requires the two $\vk$-meshes to be the same, i.e., $\mathcal{K}_i=\mathcal{K}_j=\mathcal{K}$.
The Madelung constant correction in \eqref{eqn:exchange_madelung} to the regular exchange energy calculation can also be generalized accordingly \cite{xing_unified_2023} for the case with two general  
uniform meshes $\mathcal{K}_i$ and $\mathcal{K}_j$. 

\begin{figure}[htbp]
    \centering
    \includegraphics[width=0.7\textwidth]{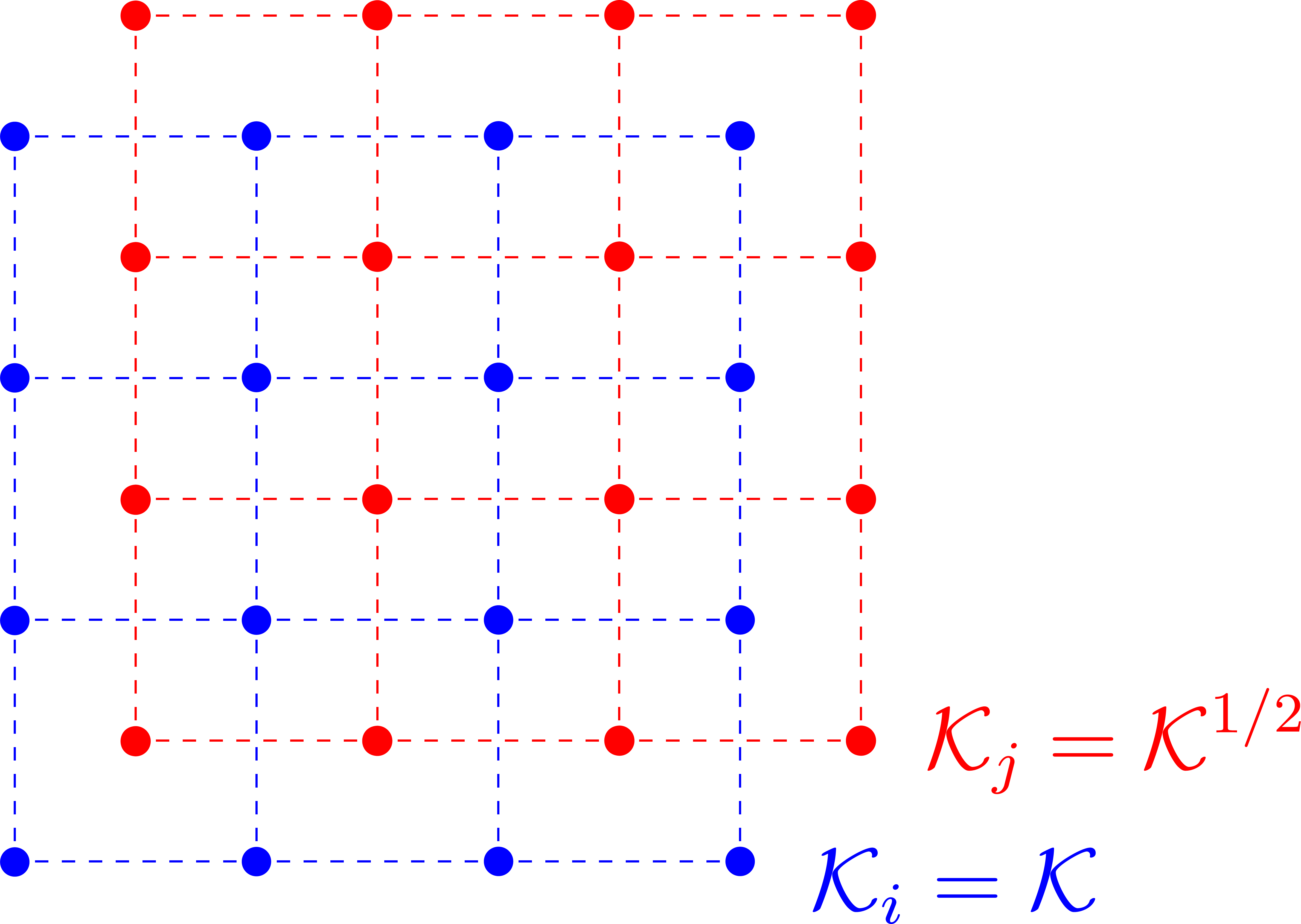}
    \caption{Schematic of the k-point grids used in the staggered mesh method in 2D.}
    \label{fig:stag-mesh-2d}
\end{figure}

The staggered mesh method for exchange energy calculation is based on the key observation that the ERI $\left( j_{\mathbf{k}_{j}}i_{\mathbf{k}_{i}}\mid i_{\mathbf{k}_{i}}j_{\mathbf{k}_{j}} \right)$ with zero momentum 
transfer $\vq = \vk_i - \vk_j = \mathbf{0}$ can lead to significant finite-size error due to the involved Coulomb singularity at $\vq + \vG = \mathbf{0}$, both with and without the Madelung constant correction.
Part of this error in a regular exchange energy calculation can be traced back to the fact that the same $\vk$-mesh is used for both $\vk_i$ and $\vk_j$. 
To eliminate this error source, the staggered mesh method applies two different meshes $\mathcal{K}_i$ and $\mathcal{K}_j$ in \eqref{eq:ek_mo} so that there is no ERI with zero momentum transfer in the exchange energy calculation.  
More specifically, we choose $\mathcal{K}_i$ to be an arbitrary uniform mesh $\mathcal{K}$ and define $\mathcal{K}_j$ as $\mathcal{K}^{1/2}$ by shifting $\mathcal{K}$ by a half-mesh-size in all directions. 
See \Cref{fig:stag-mesh-2d} for an illustration of such two $\vk$-meshes.

Due to the two staggered $\vk$-meshes $\vk_i \in \mathcal{K}$ and $\vk_j\in\mathcal{K}^{1/2}$, the minimum images of all possible momentum transfer vectors $\vq = \vk_i - \vk_j$ in $\Omega^*$ no longer lie in $\mathcal{K}_\vq$ in \eqref{eq:madelung-regular} that contains the $\Gamma$ point. 
For arbitrary $\mathcal{K}$, those momentum transfer vectors instead form another uniform mesh $\mathcal{K}_\vq^{1/2}$ which comes from a half-mesh-size shift of $\mathcal{K}_\vq$ in all directions. 
From this observation and the numerical quadrature explanation about Madelung constant correction for regular exchange energy calculation, 
the Madelung-like correction for the staggered mesh method can be generalized accordingly as (Remark 5.2 in Ref.~\citenum{xing_unified_2023})
\begin{align}
    \xi_{1/2}
    &=
    \lim_{\eta \rightarrow +\infty}\left(\frac{|\Omega^*|}{N_{\mathbf{k}}}\sum_{\mathbf{q}\in\mathcal{K}_{\mathbf{q}}^{1/2}}-\int_{\Omega^{*}}d\mathbf{q}\right)\frac{1}{(2\pi)^{3}}\sum_{\mathbf{G}\in\mathbf{L}^{*}}\frac{4\pi e^{-|\mathbf{q}+\mathbf{G}|^{2}/\eta}}{|\mathbf{q}+\mathbf{G}|^{2}}.
    \label{eq:madelung-stagger-ss}
\end{align}
This generalization consists of the first two terms in the bracket in \eqref{eqn:exchange_madelung} with $\mathcal{K}_\vq$ replaced by $\mathcal{K}_\vq^{1/2}$ while dropping the remaining two terms 
in \eqref{eqn:exchange_madelung} (these two terms decay to zero with $\eta \rightarrow+\infty$). 
Note that the above term in \eqref{eq:madelung-stagger-ss} with any $\eta > 0$ can asymptotically reduce the finite-size error and taking $\eta$ to $\infty$ is mainly to be consistent with the standard 
Madelung constant correction. 
This correction $\xi_{1/2}$ can be numerically computed by increasing $\eta$, which converges rapidly. 

Finally, the staggered mesh method with Madelung-like correction for Fock exchange energy is defined as 
\begin{equation}
    E_K^{N_\textbf{k},\xi_{1/2}}\left(\mathcal{K},\mathcal{K}^{1/2}\right)=E_K^{N_\textbf{k}}\left(\mathcal{K},\mathcal{K}^{1/2}\right)+n_{\mathrm{occ}}\xi_{1/2}.
\label{eq:ek_with_madelung}
\end{equation}
This approach to calculating the exchange energy not only reduces the errors that arise from the Coulomb singularity term in \eqref{eq:eri-reciprocal} but also 
further improves the convergence to $\mathcal{O}(N_\textbf{k}^{-5/3})$ under certain conditions \cite{xing_unified_2023}. 
Specifically, $\mathcal{O}(N_\textbf{k}^{-5/3})$ is guaranteed when the unit cell has cubic symmetry. 
In practice, as will be demonstrated next in the results, to have a cubic unit cell is a sufficient but not necessary condition.

\section{Computational Details}

\subsection{Implementation of Staggered Mesh for HF Exchange}
In practical calculations with AOs, the generalized exchange energy in \eqref{eq:ek_mo} can be formulated and computed as 
\begin{align}
E_{K}^{N_{\textbf{k}}}\left(\mathcal{K}_{i},\mathcal{K}_{j}\right)
&=-\frac{1}{2N_{\textbf{k}}}\sum_{\mathbf{k}_{i}\in\mathcal{K}_{i},\mathbf{k}_{j}\in\mathcal{K}_{j}}\sum_{\mu,\nu,\lambda,\sigma}P_{\mu\nu}^{\mathbf{k}_{j}}\left(\nu_{\mathbf{k}_{j}}\lambda_{\mathbf{k}_{i}}\mid\sigma_{\mathbf{k}_{i}}\mu_{\mathbf{k}_{j}}\right)P_{\lambda\sigma}^{\mathbf{k}_{i}}\label{eq:ek_ao}
\\
&=-\frac{1}{2N_{\textbf{k}}}\sum_{\mathbf{k}_{j}\in\mathcal{K}_{j}}\sum_{\mu,\nu}P_{\mu\nu}^{\mathbf{k}_{j}}K_{\nu\mu}^{\mathbf{k}_{j}},
\label{eq:ek_kmat}
\end{align}
where the exchange matrix is defined as 
\begin{equation}
    K_{\nu\mu}^{\mathbf{k}_{j}}=\frac{\partial E_{K}}{\partial P_{\mu\nu}^{\mathbf{k}_{j}}}=-\sum_{\mathbf{k}_{i}\in \mathcal{K}_i}\sum_{\sigma\lambda}\left(\nu_{\mathbf{k}_{j}}\lambda_{\mathbf{k}_{i}}\mid\sigma_{\mathbf{k}_{i}}\mu_{\mathbf{k}_{j}}\right)P_{\lambda\sigma}^{\mathbf{k}_{i}}.
    \label{eq:kmat}
\end{equation}

We introduce the staggered mesh method along with two computationally cheaper versions, which essentially all differ in how they compute the densities $P^{\mathbf{k}}$ in equation \eqref{eq:ek_ao}.

\begin{enumerate}
    \item \textit{(Original) staggered mesh.} This delineates the original version of the staggered mesh method.   In this procedure, an SCF calculation is done on the \textit{combined} set of the unshifted and shifted k-points. In other words, this means solving for $P^{\mathbf{k}_l}$, where $\mathbf{k}_l\in \mathcal{K}^{0+1/2}= \mathcal{K}\cup \mathcal{K}^{1/2}$. If the original unshifted \textbf{k}-mesh has $N_\textbf{k}$ \textbf{k}-points, then this SCF calculation solves for a total of $2N_\textbf{k}$ \textbf{k}-points. Then, we construct $K^{\mathbf{k}_j}$ via \eqref{eq:kmat}, where $\mathbf{k}_j\in\mathcal{K}_j=\mathcal{K}^{1/2}$ and $\mathbf{k}_i\in\mathcal{K}_i=\mathcal{K}$. Finally, we compute the corrected exchange energy via \eqref{eq:ek_kmat} and \eqref{eq:ek_with_madelung},  As the construction of $K$ dominates the SCF calculation cost with a scaling of $\mathcal{O}(N_\textbf{k}^2)$, this approach is expected to cost approximately 4x as much as a regular HF exchange calculation, making it the most expensive of the three options described here. 
    To keep the subsequent $N_\textbf{k}$ convergence plots consistent with the other methods, we say that this version of staggered mesh uses $N_\mathbf{k}$ points
    if the unshifted mesh $\mathcal{K}$ \textit{by itself} contains $N_\mathbf{k}$ points.  

    \item \textit{Split-SCF staggered mesh.} In terms of computational expense, an intermediate option is to follow the approach above, 
    except we perform two separate SCF calculations, one on $\mathcal{K}$ and the other on $\mathcal{K}^{1/2}$. 
    This allows us to separately compute $P_{\mu\nu}^{\mathbf{k}_i}$ and $P_{\mu\nu}^{\mathbf{k}_j}$. This is then followed by the exchange energy computation in \eqref{eq:ek_kmat}. Therefore, the cost is expected to be 2x that of an SCF calculation. Here, there will still be some discrepancies arising from the fact that the two densities are effectively being optimized in different mean-field potentials.
      
    \item \textit{Non-SCF staggered mesh.} This is the cheapest of the three options. First, an SCF calculation is done on  $\mathcal{K}$, so $\mathbf{k}_i,\mathbf{k}_j\in \mathcal{K}_i=\mathcal{K}$, which gives us $P^{\mathbf{k}_{i}}$. Then, using this density matrix to create an effective potential, we construct $K^{\mathbf{k}_j}$ via \eqref{eq:kmat}, where $\mathbf{k}_i\in \mathcal{K}_i=\mathcal{K}$, but $\mathbf{k}_j\in\mathcal{K}_j=\mathcal{K}^{1/2}$. This exchange matrix is used to build the Fock matrix $F^{\mathbf{k}_j}$ which is then diagonalized to yield $P^{\mathbf{k}_{j}}$, that is, the elements of the density matrix on the shifted \textbf{k}-mesh. This is equivalent to performing a band structure calculation on the shifted \textbf{k}-mesh based on the mean-field potential induced by the densities at the unshifted \textbf{k}-mesh. Finally, this new density matrix is used to calculate the staggered mesh exchange energy in \eqref{eq:ek_kmat}.


\end{enumerate}

The staggered mesh method, along with the Split and Non-SCF versions,
are implemented in a development version of Q-Chem \cite{epifanovsky_software_2021}. 
Unless specified otherwise, all geometries were obtained from the Materials Project\cite{jain_high-throughput_2011}, which were optimized using the PBE functional \cite{perdew_generalized_1996}.
The exceptions are the ammonia crystal, which was obtained from the X23 dataset \cite{dolgonos_revised_2019} and the \ce{H2} molecule, which is two H atoms at the center of a 6-bohr box with an interatomic distance of 1.4 bohr. All calculations were performed using the SZV-GTH basis set \cite{vandevondele_gaussian_2007} except for the band gap calculations, which were performed with the uncontracted def2-QZVP-GTH basis set\cite{lee_approaching_2021}. For computing pair densities in the two-electron integrals, we used the Gaussian and Planewaves (GPW) density fitting scheme, also known as Fast-Fourier-Transform Density Fitting (FFTDF) 
 \cite{lippert_hybrid_1997,vandevondele_gaussian_2007}. 

\section{Results}

\subsection{Exchange Energies}
A total of twelve crystal systems were tested for the convergence of the exact exchange energy $E_{K}$. Among these systems, two are molecular (\ce{H2} and ammonia), and the rest are simple solids made of either one (diamond, silicon, graphite) or two elements (LiH; LiF; SiC; hexagonal and cubic BN; AlN; and MgO). Plotting the computed exchange energies versus increasing number of $\mathbf{k}$-points is shown in \cref{fig:all_nk_convergence}. For some systems (especially those with large unit cells and hence a large $n_{AO}$), certain lines and data points are missing due to insufficient compute capabilities. In particular, for AlN, the Stagger and Stagger Split-SCF lines stop at $N_{\mathbf{k}}=4^3$ and $N_{\mathbf{k}}=5^3$ respectively, and for the ammonia crystal, these lines are not available. 

The trend that stands out in these results is that all staggered mesh methods display accelerated convergence towards the TDL, even outperforming the Truncated Coulomb method. This is a significant result given that the Truncated Coulomb kernel converges exponentially towards the TDL for insulating systems due to the exponential localization of the single-particle density matrix.\cite{spencer_efficient_2008,kohn_density_1996}
Amongst the three variants of the staggered mesh method, the original version performs the best (albeit by a small margin) for most systems as expected, with the notable exceptions of SiC, Si, Cubic BN, and AlN where the Non-SCF and Split-SCF versions perform slightly better. 

However, the original version of the staggered mesh method has a larger memory footprint. For large systems such as the ammonia crystal system and AlN, we can only perform the Non-SCF version of the staggered mesh method. 


\begin{figure}
     \centering
     \begin{subfigure}[b]{0.32\textwidth}
         \centering
         \includegraphics[width=\textwidth]{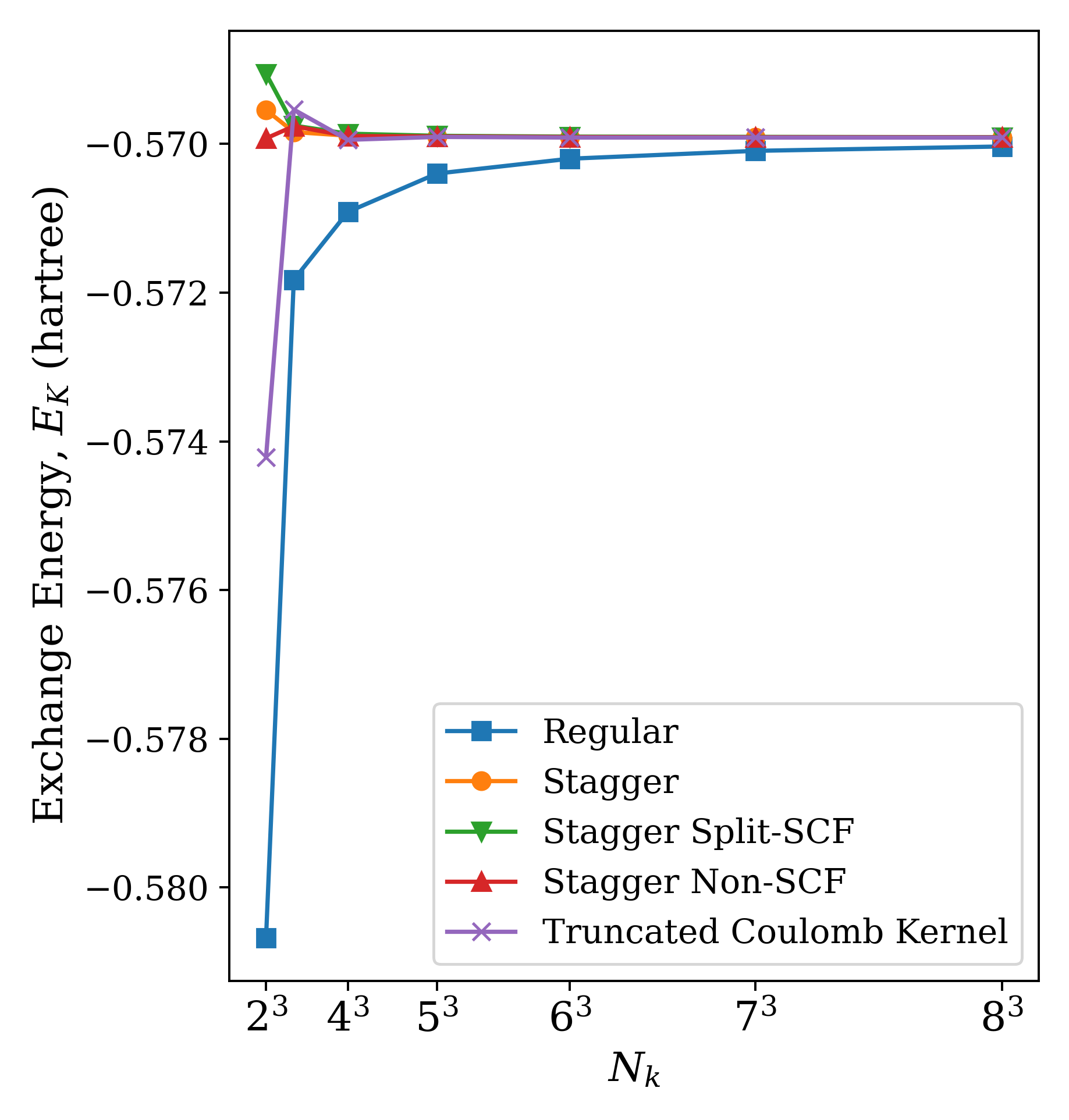}
         \caption{\ce{H2}}
         \label{fig:H2_nk8-512_general}
     \end{subfigure}
     \hfill
     \begin{subfigure}[b]{0.32\textwidth}
         \centering
         \includegraphics[width=\textwidth]{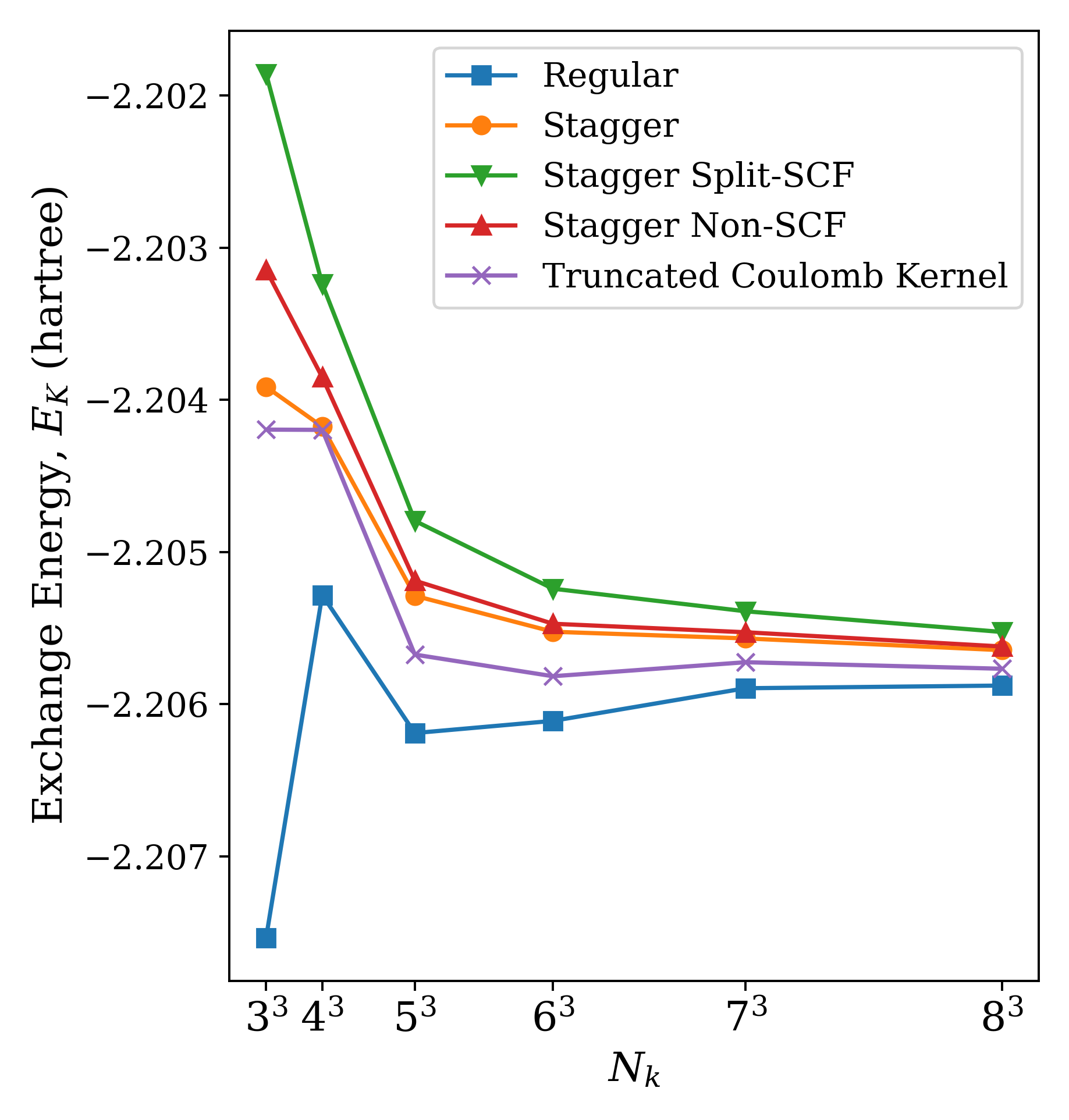}
         \caption{\ce{LiH}}
         \label{fig:general_nk/LiH_nk27-512_general}
     \end{subfigure}
     \hfill
     \begin{subfigure}[b]{0.32\textwidth}
         \centering
         \includegraphics[width=\textwidth]{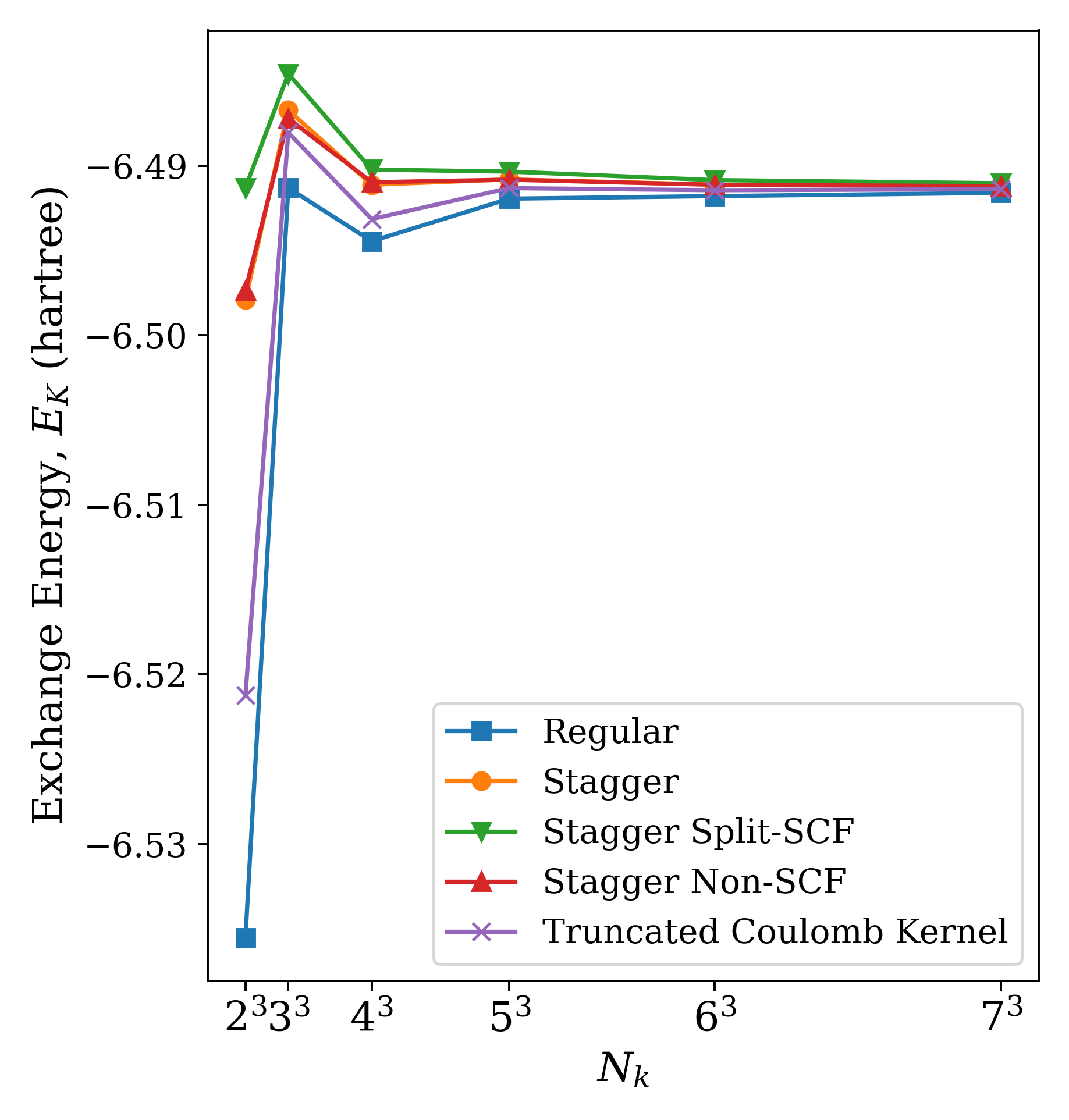}
         \caption{LiF}
         \label{fig:LiF_nk8-512_general}
     \end{subfigure}

     \begin{subfigure}[b]{0.32\textwidth}
         \centering
         \includegraphics[width=\textwidth]{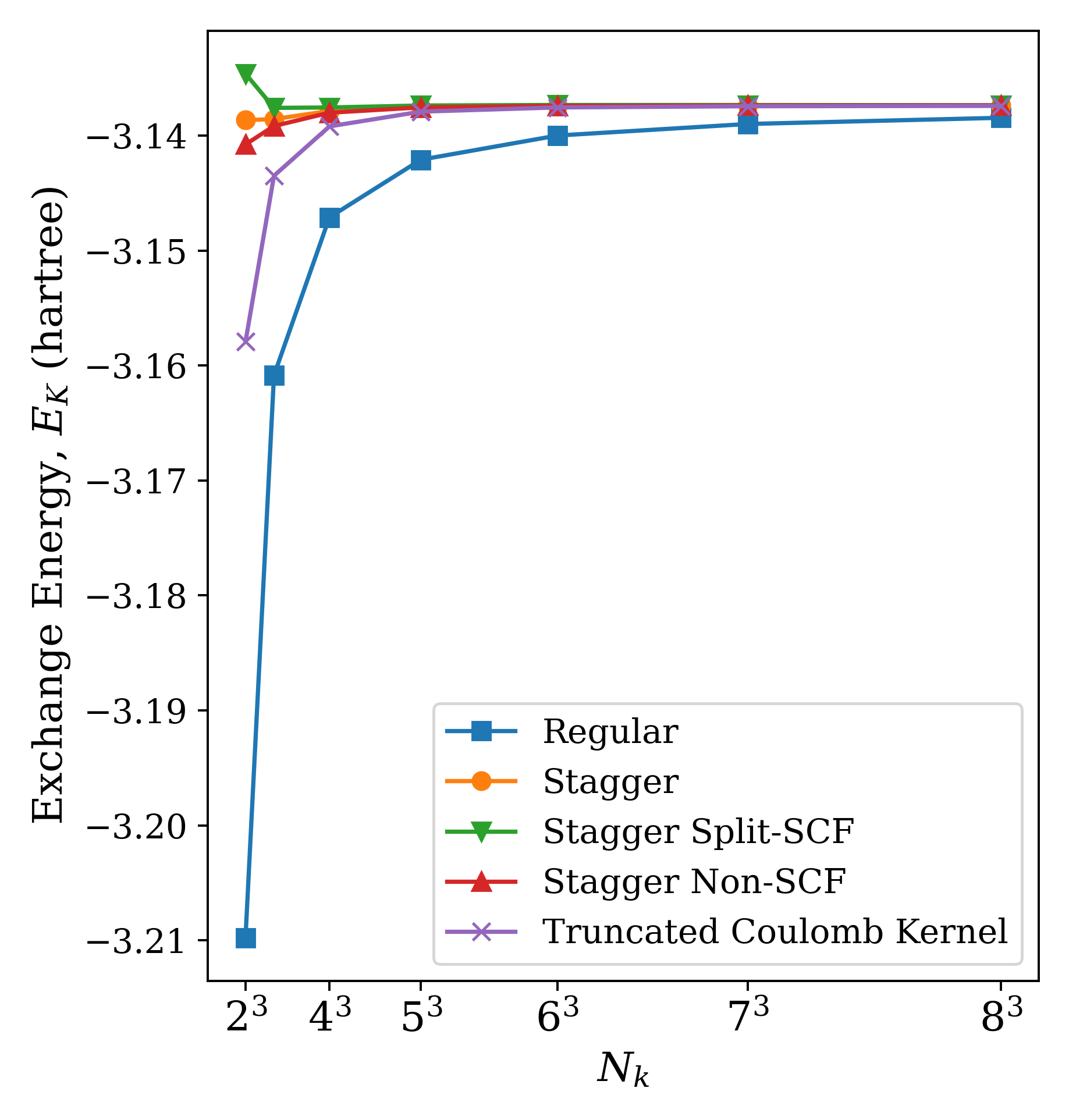}
         \caption{Diamond}
         \label{fig:diamond_mp66_nk8-512_general}
     \end{subfigure}
     \hfill
     \begin{subfigure}[b]{0.32\textwidth}
         \centering
         \includegraphics[width=\textwidth]{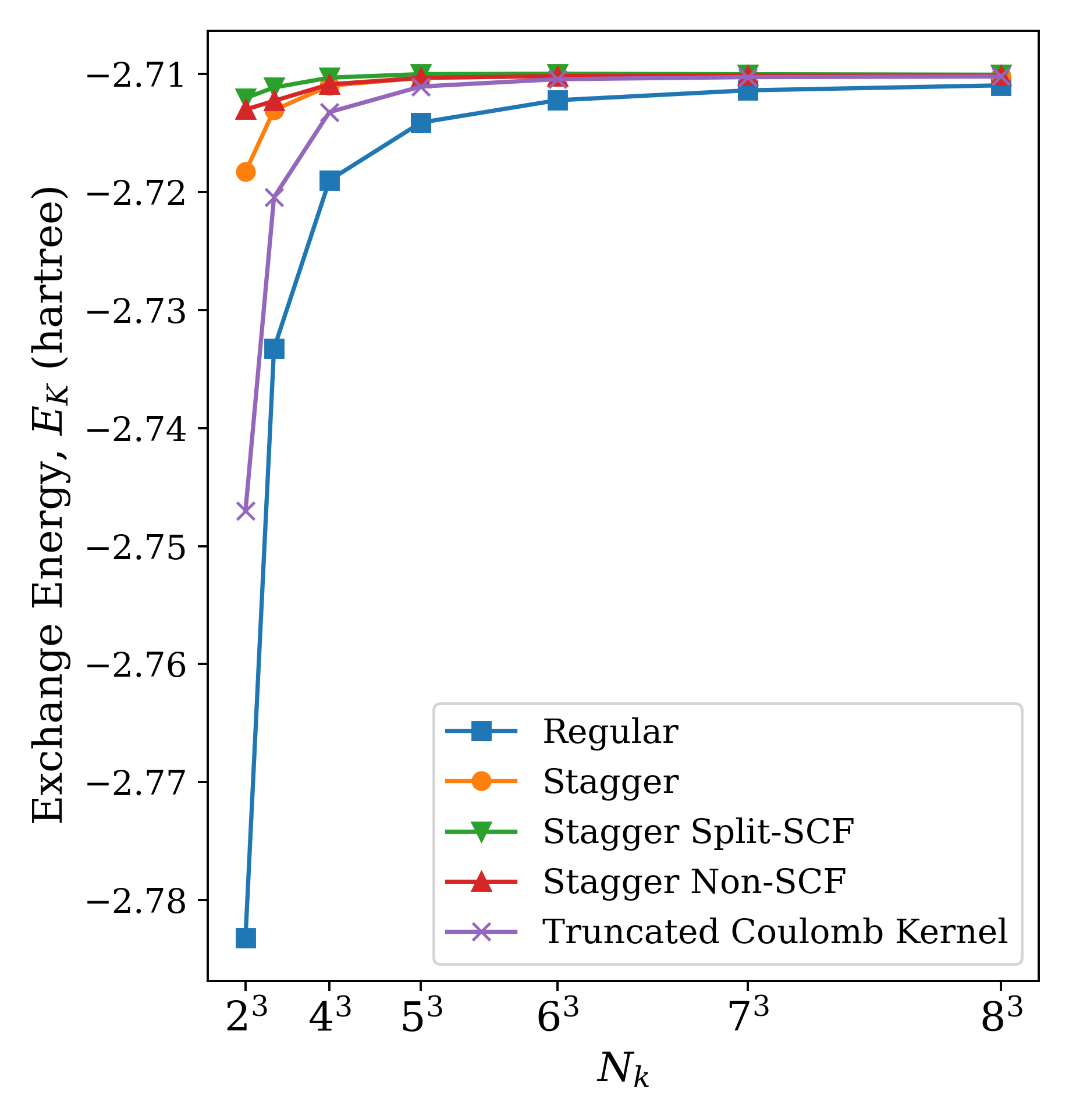}
         \caption{SiC}
         \label{fig:SiC_mp8062_nk8-512_general}
     \end{subfigure}
     \hfill
     \begin{subfigure}[b]{0.32\textwidth}
         \centering
         \includegraphics[width=\textwidth]{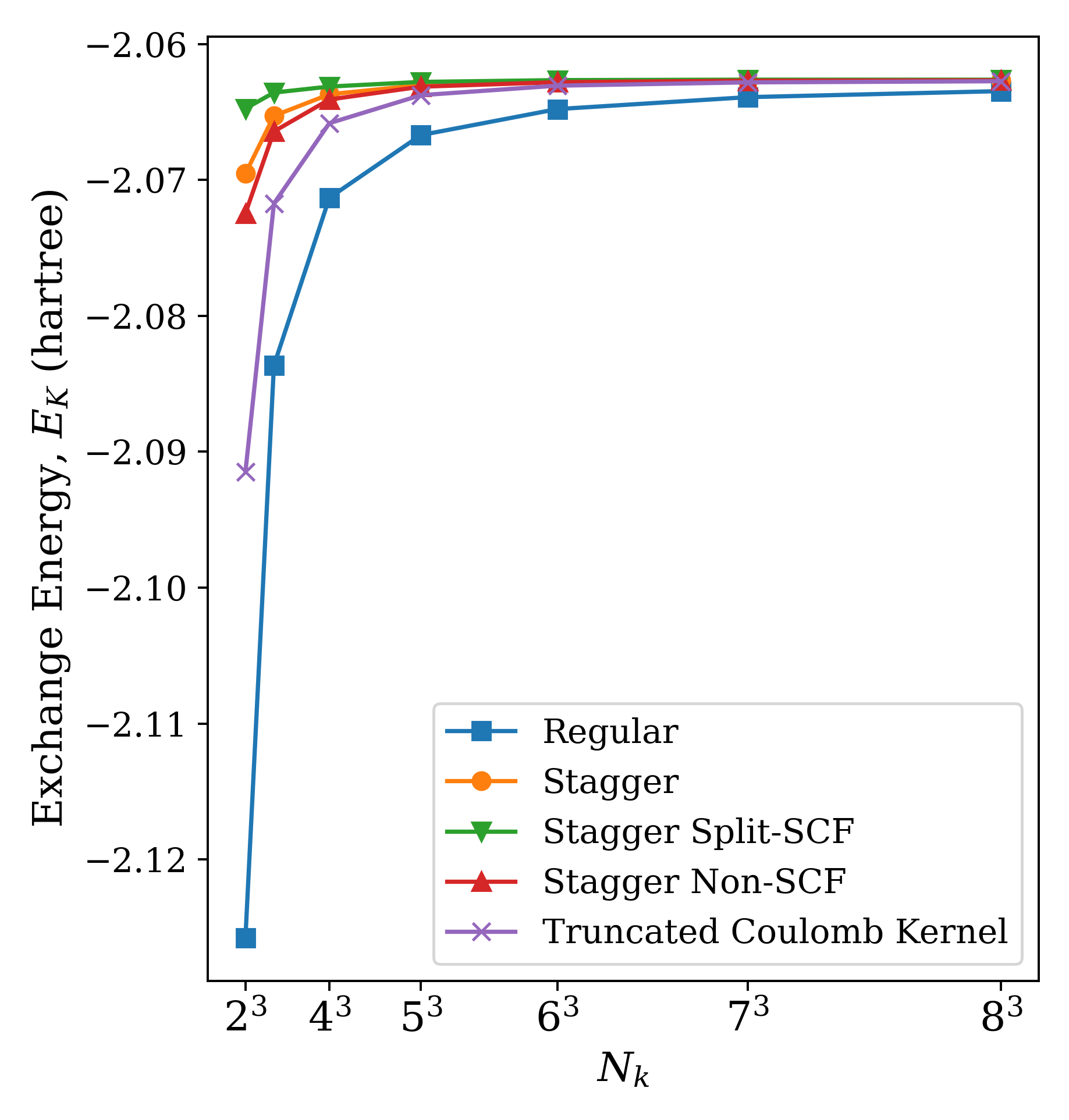}
         \caption{Si}
         \label{fig:Si_mp149_nk8-512_general}
     \end{subfigure}

     \begin{subfigure}[b]{0.32\textwidth}
         \centering
         \includegraphics[width=\textwidth]{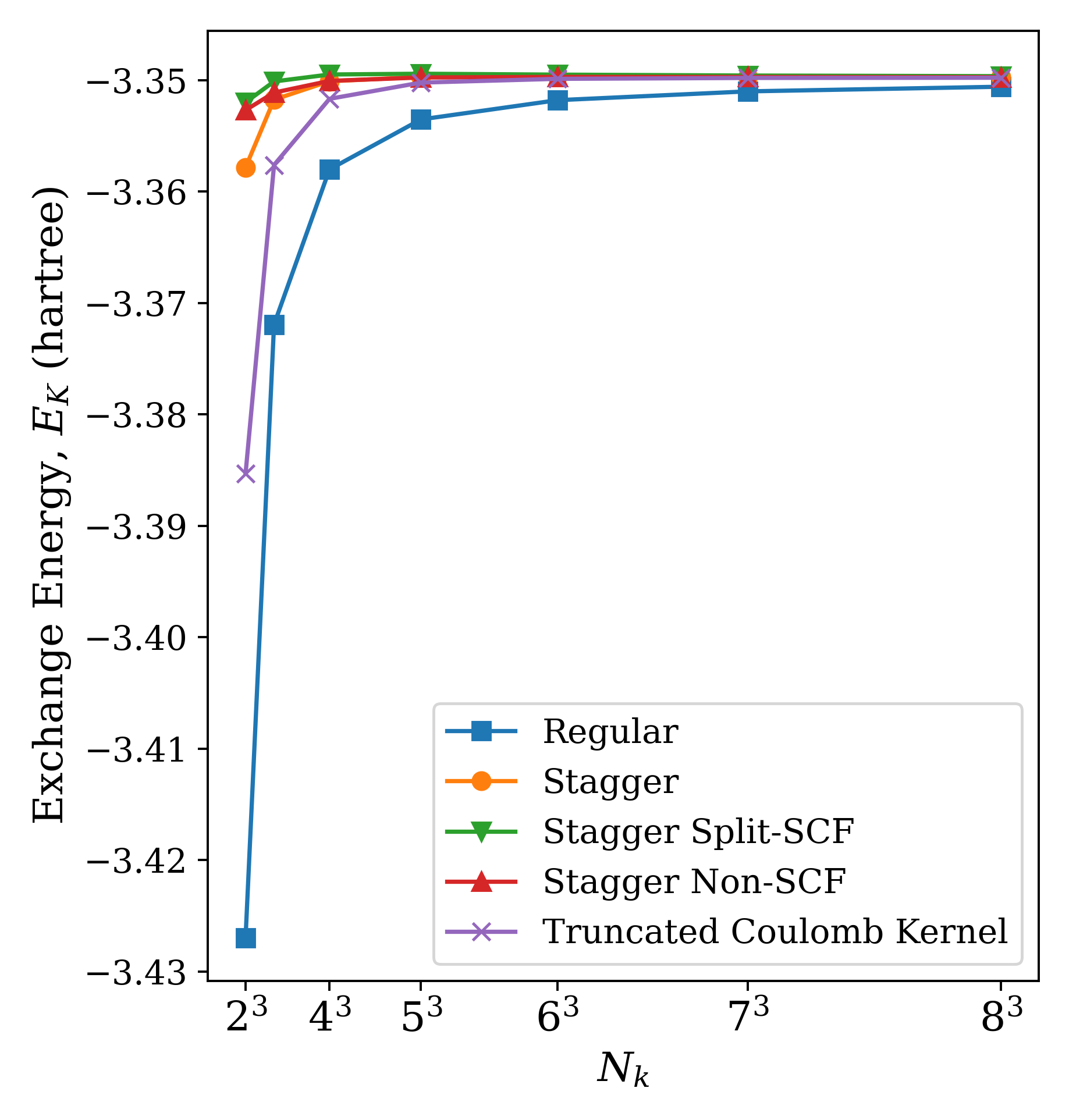}
         \caption{Cubic BN}
         \label{fig:BN_mp1639_nk8-512_general}
     \end{subfigure}
     \hfill
     \begin{subfigure}[b]{0.32\textwidth}
         \centering
         \includegraphics[width=\textwidth]{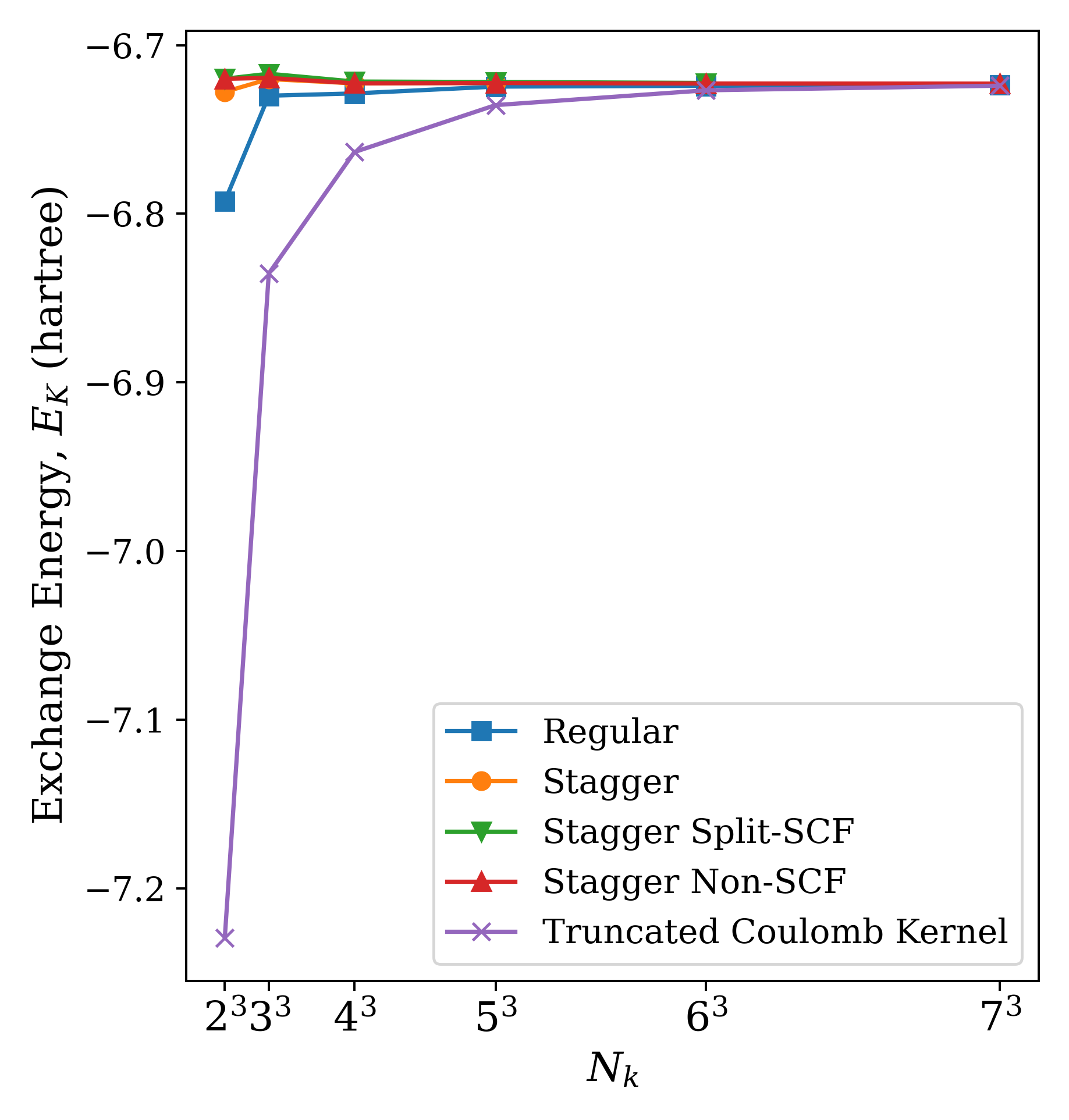}
         \caption{Hexagonal BN}
         \label{fig:BN_hex_nk8-512_general}
     \end{subfigure}
     \hfill
     \begin{subfigure}[b]{0.32\textwidth}
         \centering
         \includegraphics[width=\textwidth]{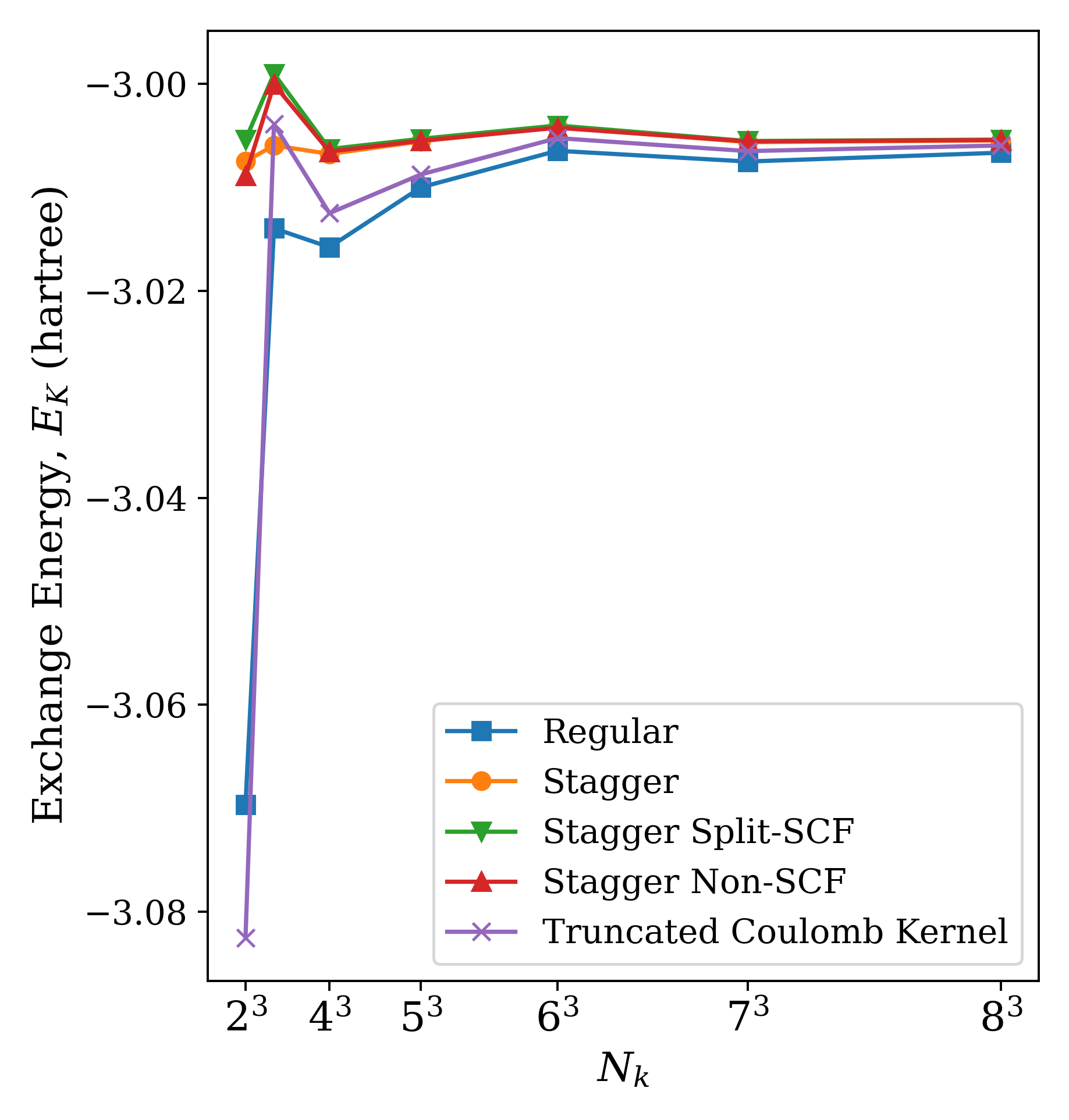}
         \caption{Graphite}
         \label{fig:graphite_nk8-512_general}
     \end{subfigure}

\end{figure}

\begin{figure}\ContinuedFloat
     \centering
     \begin{subfigure}[b]{0.32\textwidth}
         \centering
         \includegraphics[width=\textwidth]{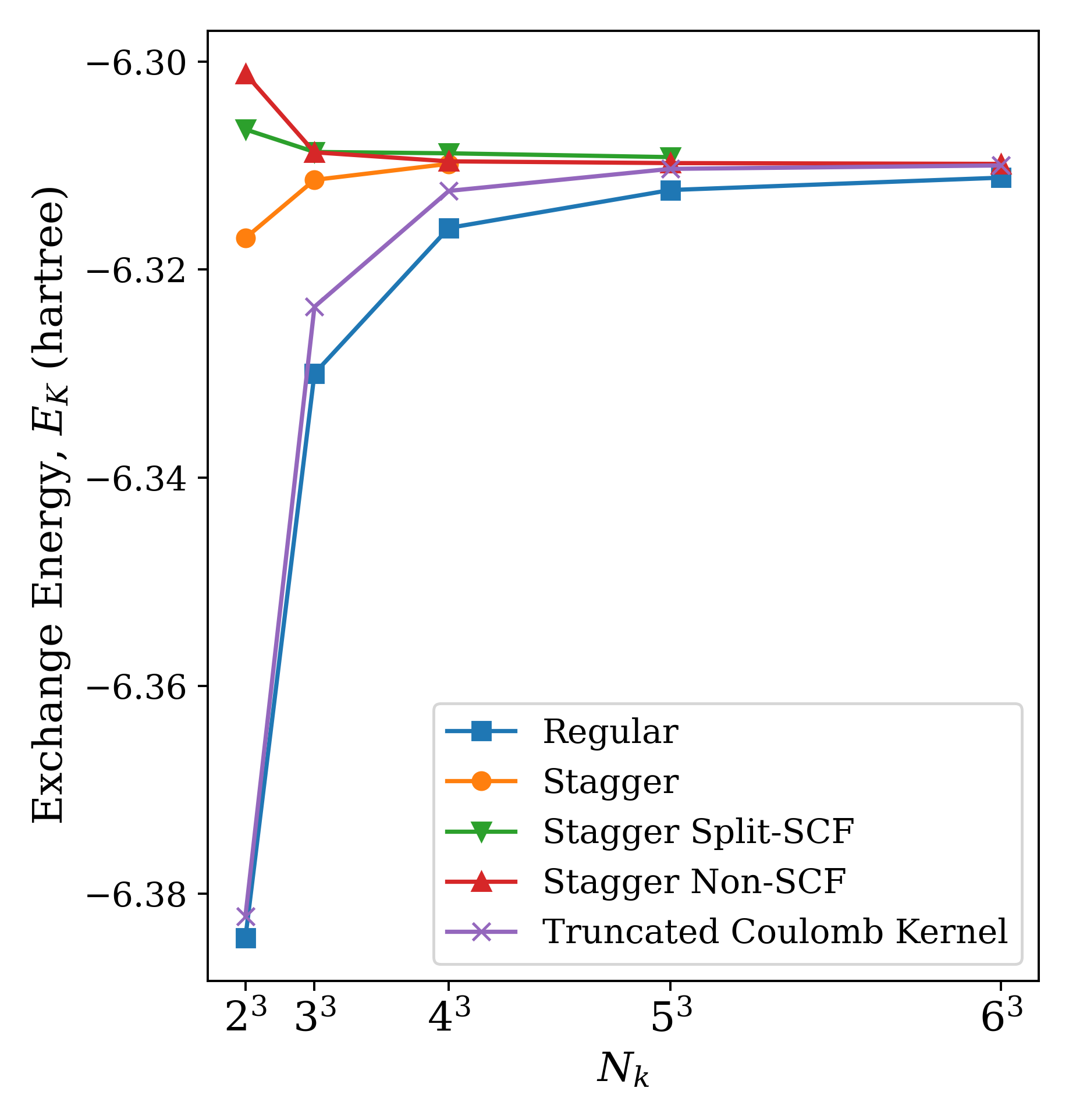}
         \caption{AlN}
         \label{fig:AlN_nk8-512_general}
     \end{subfigure}
     \hfill
     \begin{subfigure}[b]{0.32\textwidth}
         \centering
         \includegraphics[width=\textwidth]{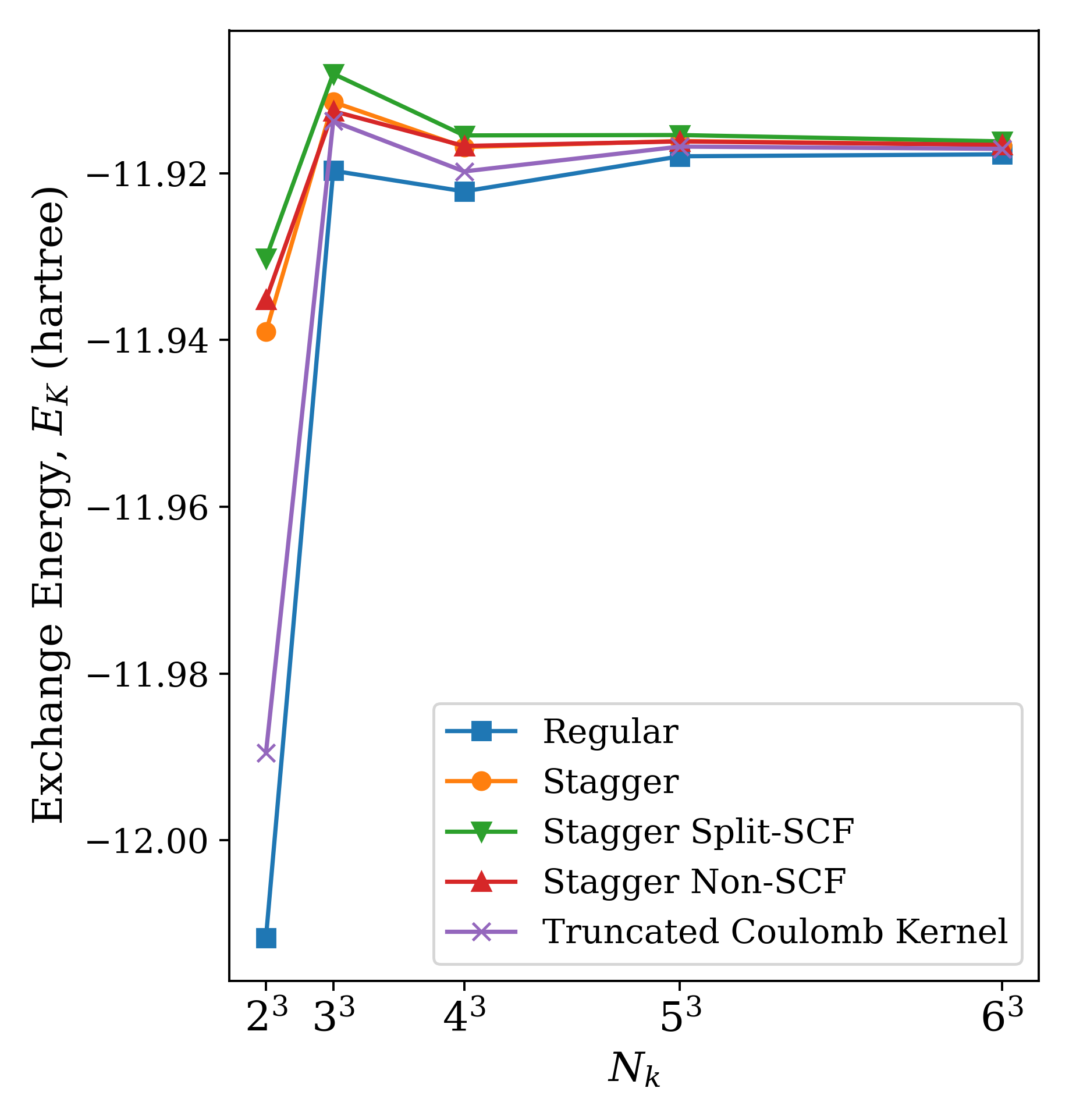}
         \caption{MgO}
         \label{fig:MgO_nk8-512_general}
     \end{subfigure}
     \hfill
     \begin{subfigure}[b]{0.32\textwidth}
         \centering
         \includegraphics[width=\textwidth]{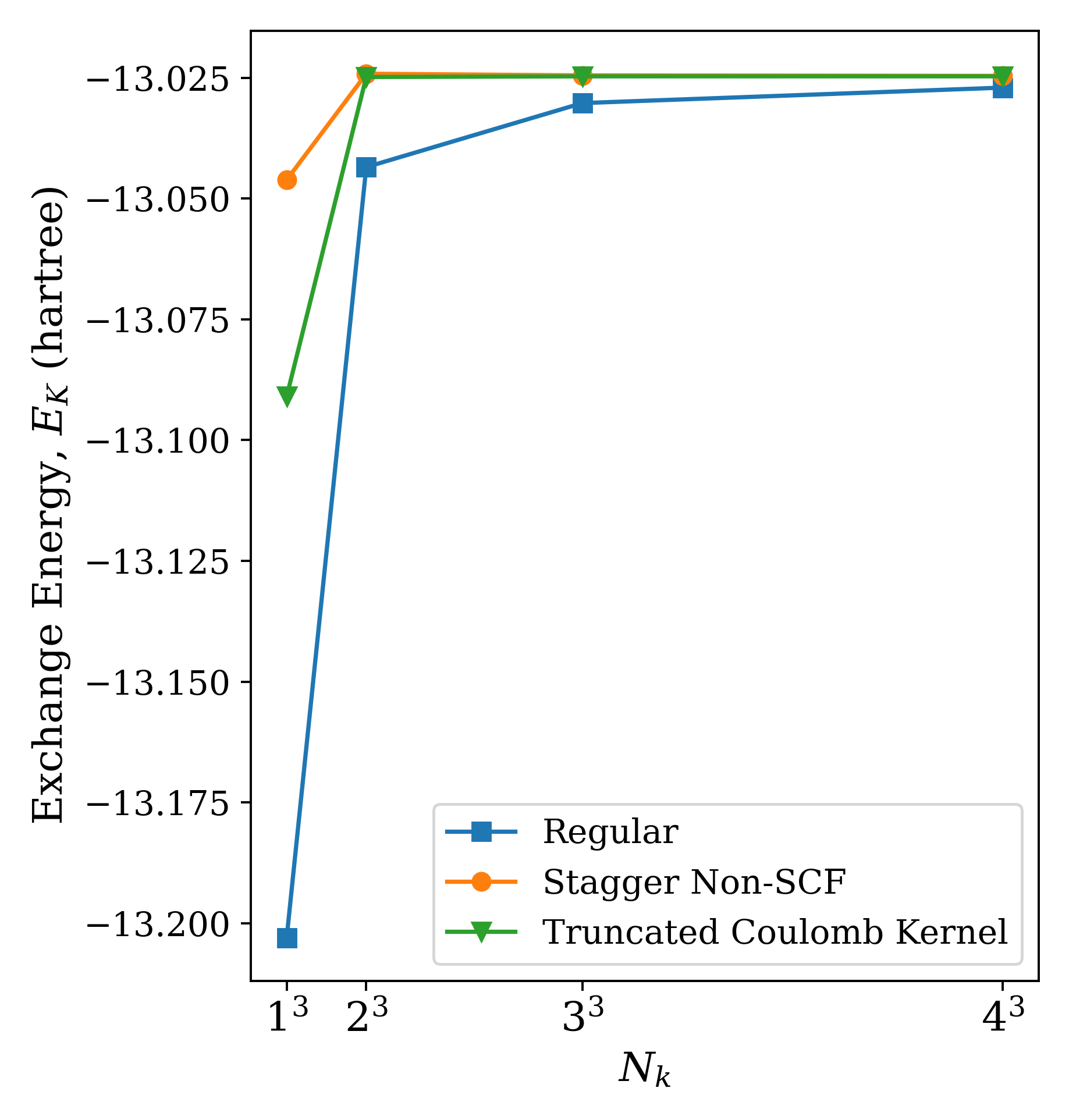}
         \caption{Ammonia Crystal}
         \label{fig:ammonia_nk1-512_general}
     \end{subfigure}

        \caption{$N_\mathbf{k}$ Convergence Plots}
        \label{fig:all_nk_convergence}

\end{figure}

\begin{figure}
     \centering
     \begin{subfigure}[b]{0.45\textwidth}
         \centering
         \includegraphics[width=\textwidth]{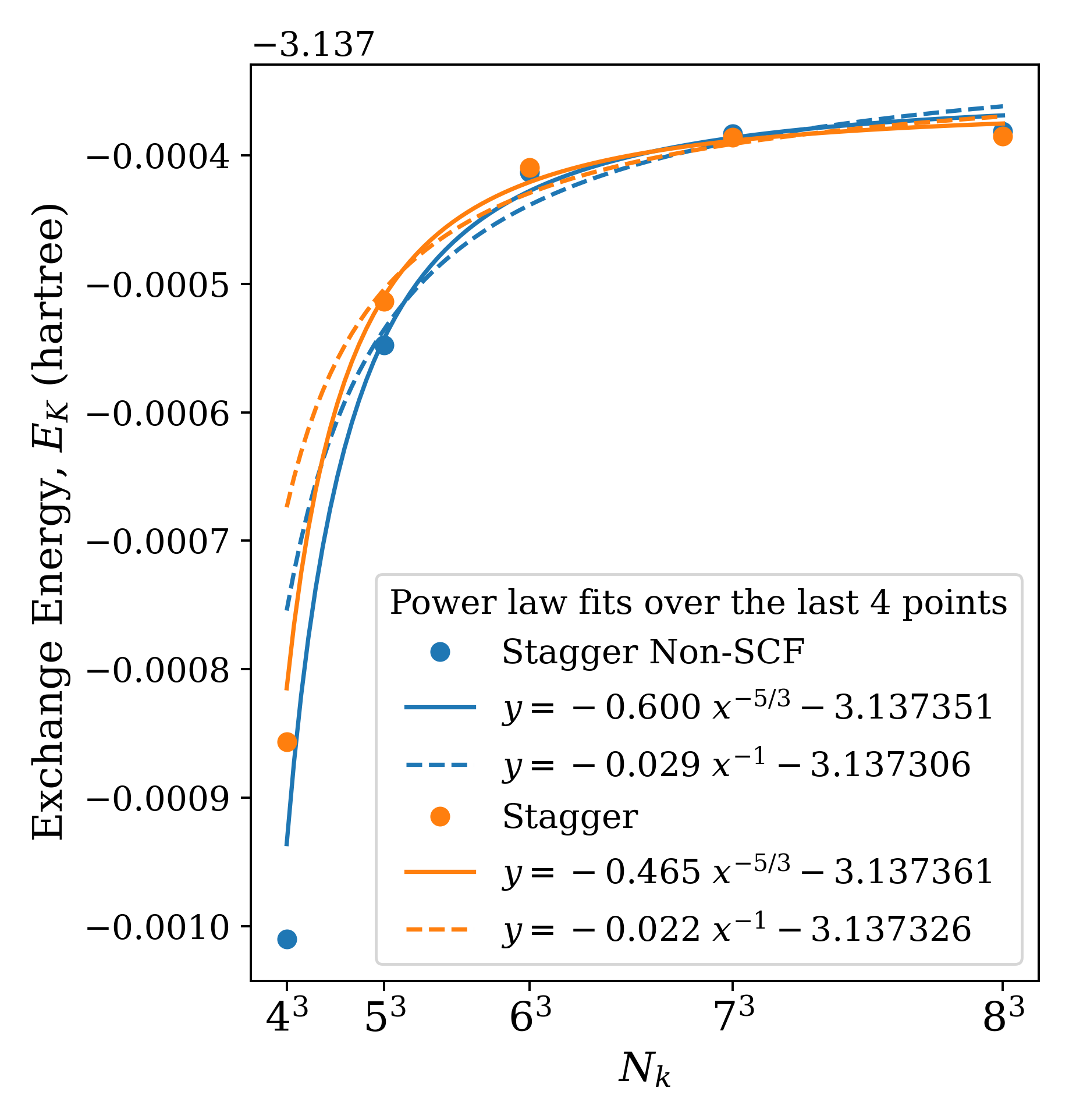}
         \caption{Diamond}
         
     \end{subfigure}
     \hfill
     \begin{subfigure}[b]{0.45\textwidth}
         \centering
         \includegraphics[width=\textwidth]{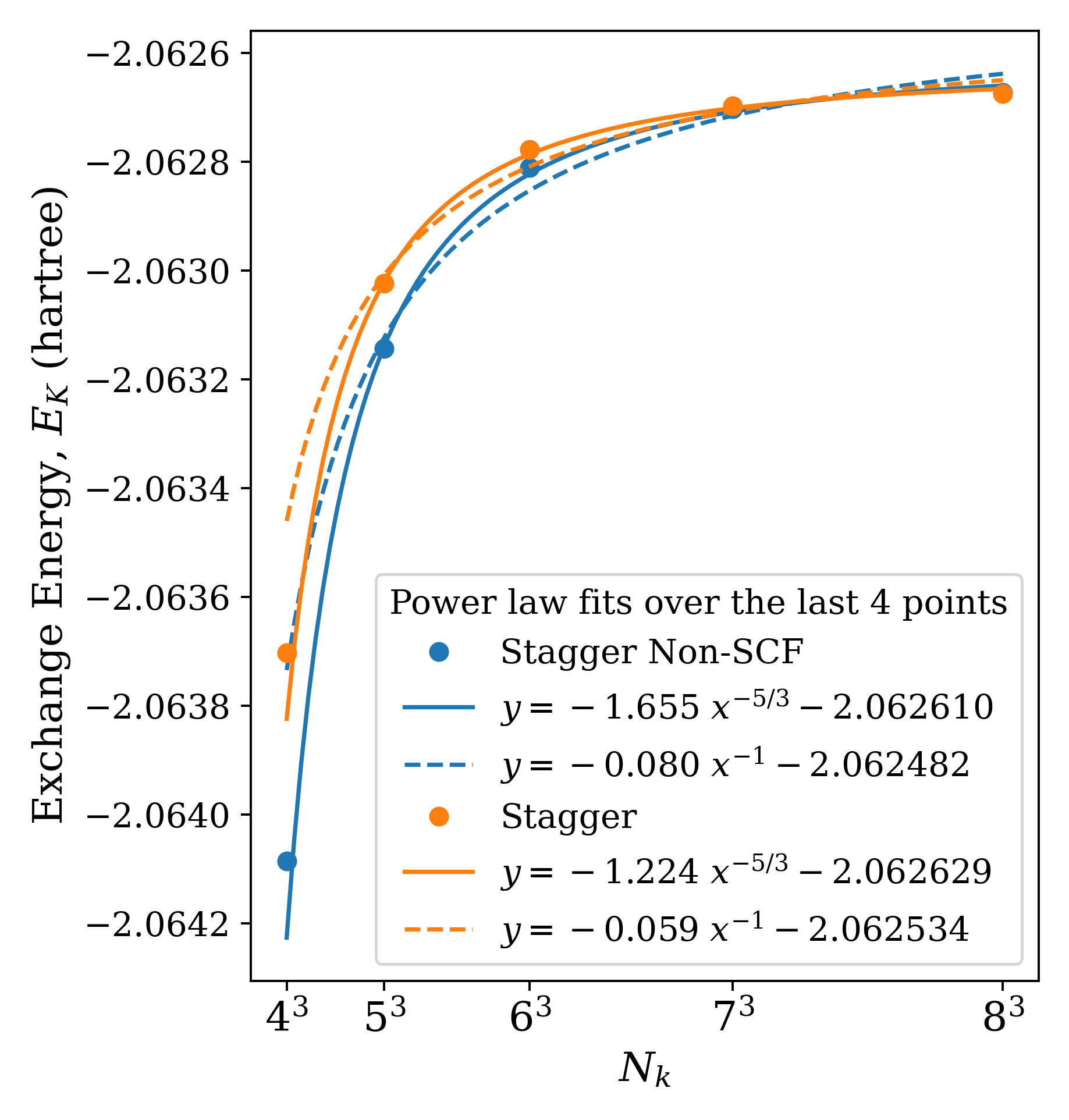}
         \caption{Si}
         \label{fig:five over x}
     \end{subfigure}
        \caption{Power Law Fits for the Exchange Energy}
        \label{fig:all_power_law_fits}
\end{figure}

\paragraph{Asymptotic Convergence}
In \cref{fig:all_power_law_fits}, a power-law curve is fitted to the convergence of energy in diamond and silicon, which both have a rhombohedral primitive cell and exhibit cubic symmetry. In addition to a more favorable prefactor than the regular method, these curves demonstrate a convergence rate faster than $\mathcal{O}(N_{\mathbf{k}}^{-1})$, aligning more closely with the $\mathcal{O}(N_{\mathbf{k}}^{-5/3})$ trend. Conversely, we can expect that the staggered mesh method could become less effective
(i.e. converge at the same rate as the regular method) in systems where the unit cell has lower symmetry\cite{xing_unified_2023}. To highlight this, we compare diamond (\cref{fig:diamond_mp66_nk8-512_general}) to graphite (\cref{fig:graphite_nk8-512_general}), which both have two carbon atoms in their primitive cells, but graphite has trigonal symmetry instead of cubic. There, we also observe that the staggered mesh method experiences approximately the same degree of oscillation as the regular and Truncated Coulomb methods even when $N_\mathbf{k}=6^3$ or higher.   

\paragraph{Basis Set Effects} We also tested the effect of using the larger GTH-DZVP AO basis set on the performance of the different versions of the staggered mesh method 
on LiH, BN, silicon, and diamond. The results, shown in \cref{fig:dzvp-results} show similar shapes to their SZV-GTH counterparts, with the three staggered mesh methods giving much more similar curves to each other than to the regular method. This points to the use of the Non-SCF version being the most cost-effective variant, 
while providing similar convergence rates to the original and Split-SCF versions.

\begin{figure}
    \centering
     \begin{subfigure}[b]{0.45\textwidth}
         \centering
         \includegraphics[width=\textwidth]{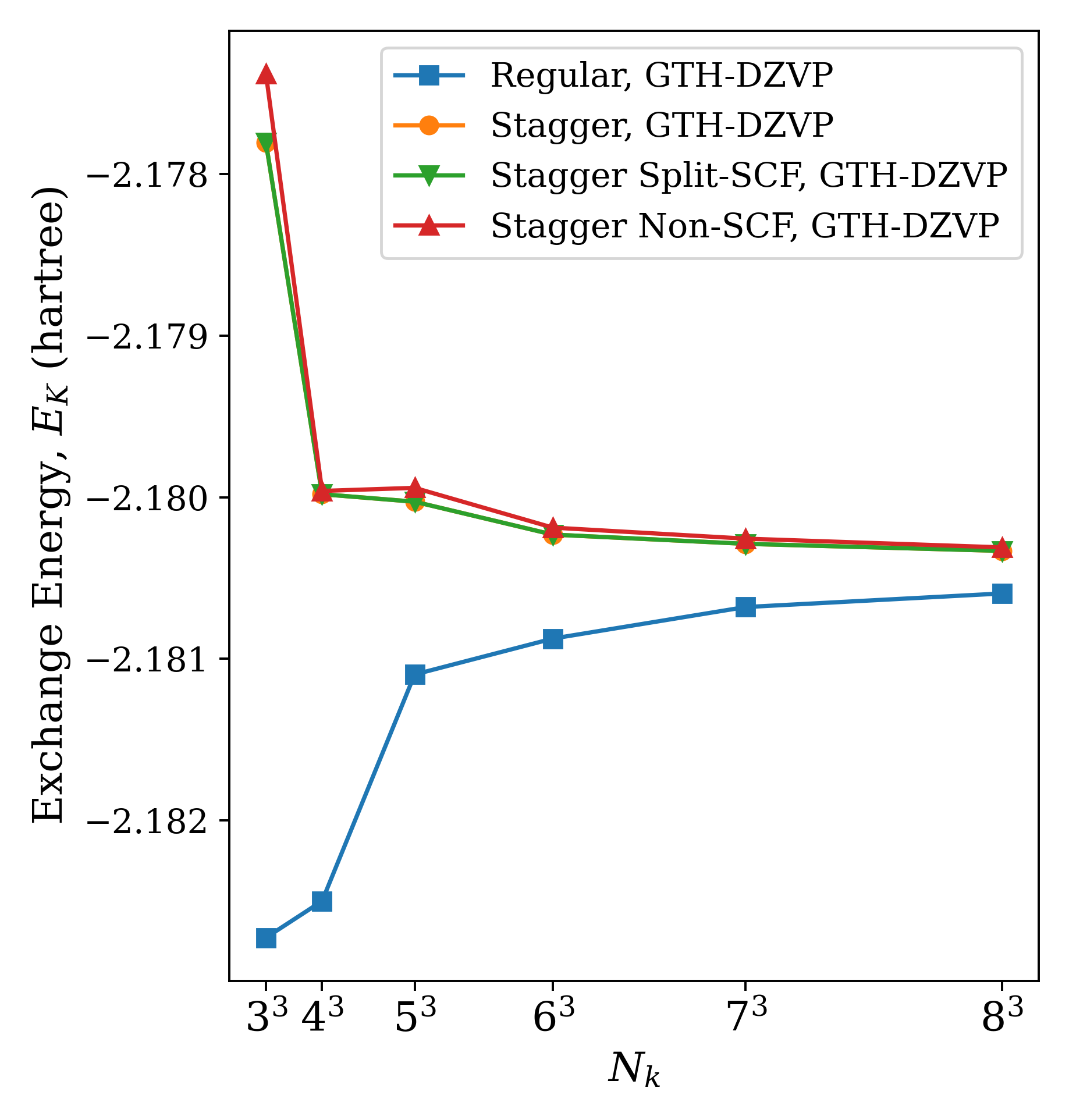}
         \caption{LiH. Compare with \cref{fig:general_nk/LiH_nk27-512_general}}
         \label{fig:lih_dzvp}
     \end{subfigure}
     \hfill
     \begin{subfigure}[b]{0.45\textwidth}
         \centering
         \includegraphics[width=\textwidth]{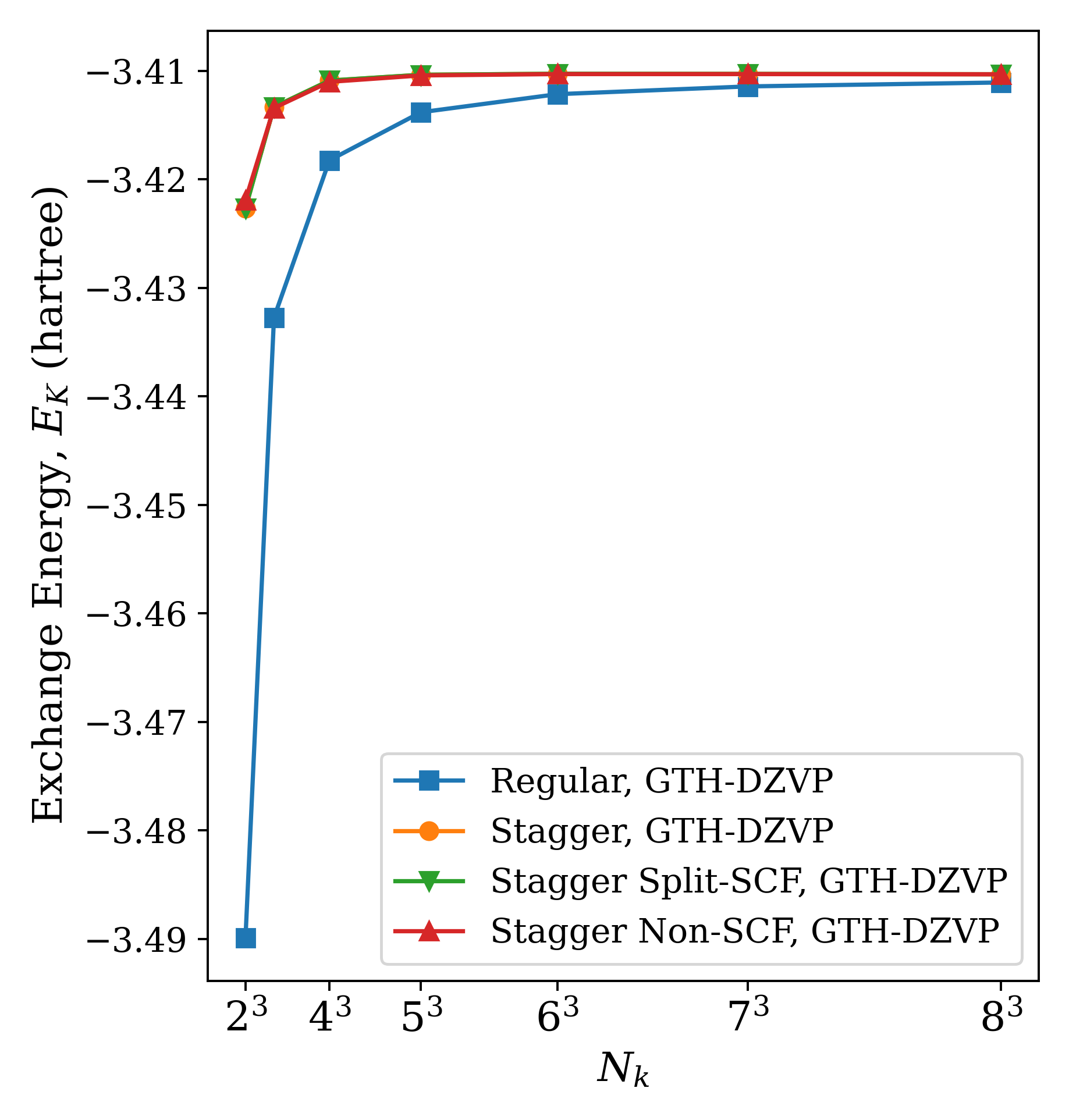}
         \caption{BN. Compare with \cref{fig:BN_mp1639_nk8-512_general}}
         \label{fig:bn_dzvp}
     \end{subfigure}
\end{figure}

\begin{figure}\ContinuedFloat
    \centering
     \begin{subfigure}[b]{0.45\textwidth}
         \centering
         \includegraphics[width=\textwidth]{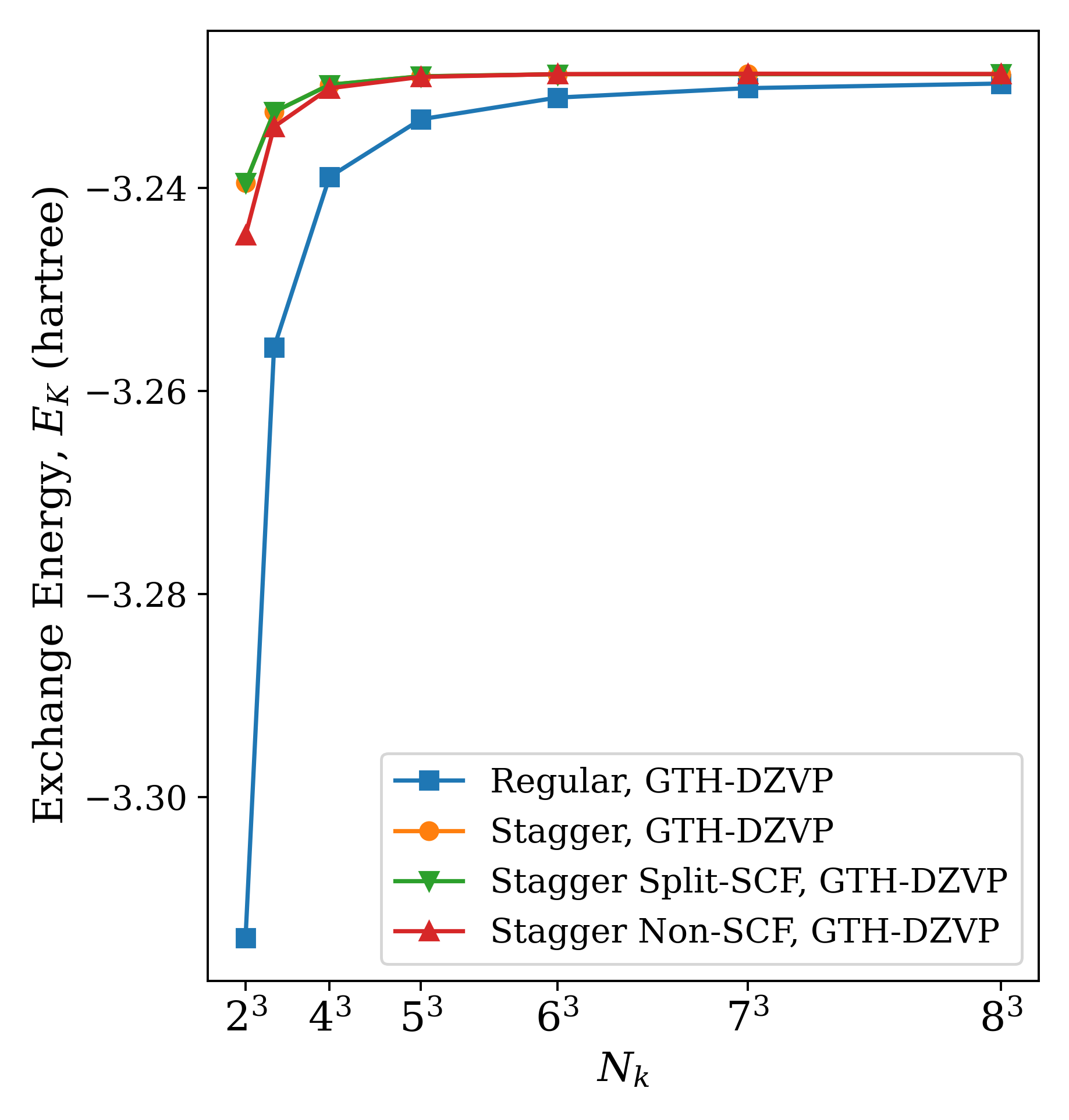}
         \caption{Diamond. Compare with \cref{fig:diamond_mp66_nk8-512_general}}
         \label{fig:diamond_dzvp}
     \end{subfigure}
     \hfill
     \begin{subfigure}[b]{0.45\textwidth}
         \centering
         \includegraphics[width=\textwidth]{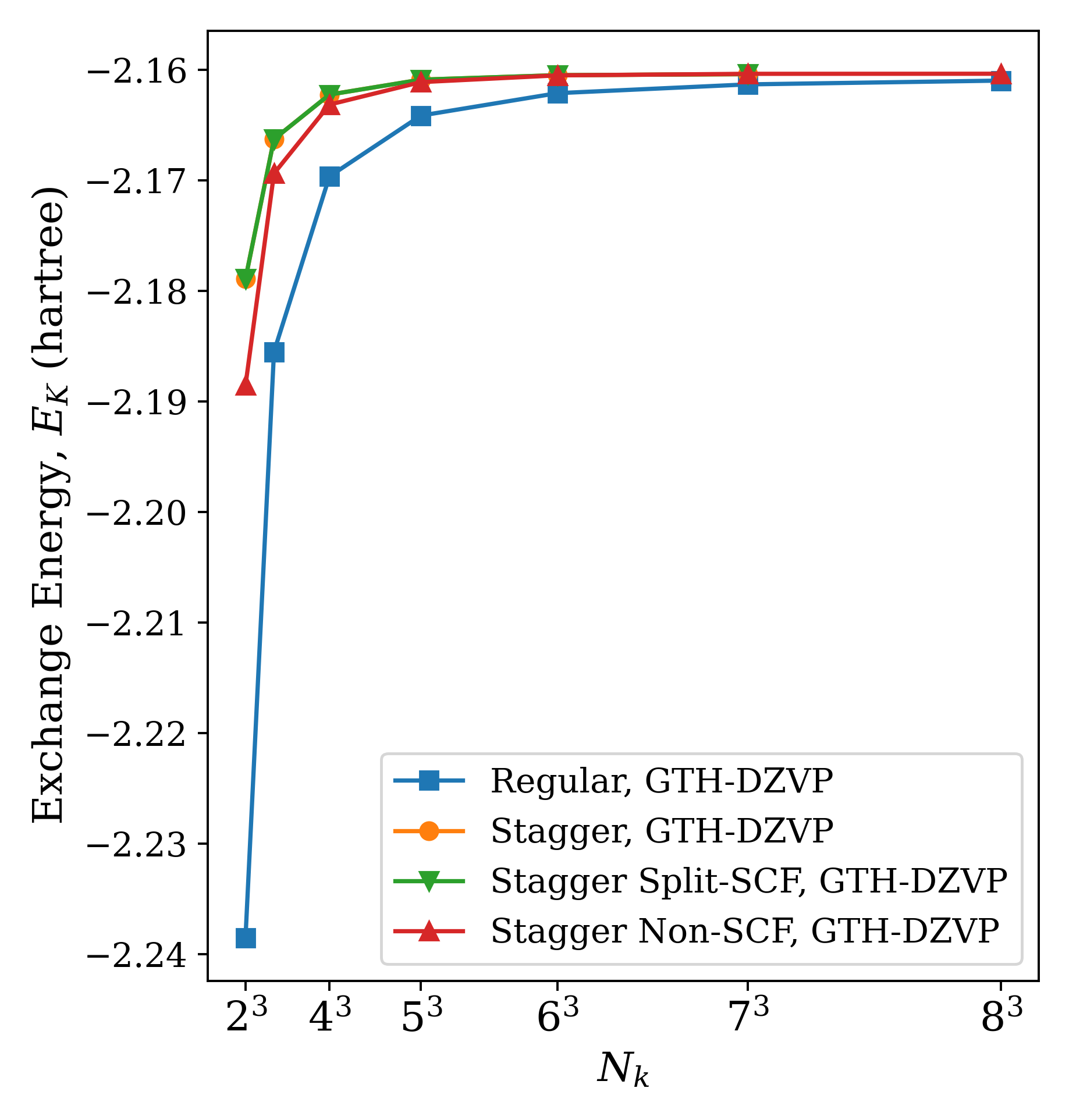}
         \caption{Silicon. Compare with \cref{fig:Si_mp149_nk8-512_general}}
         \label{fig:si_dzvp}
     \end{subfigure}
    \caption{Finite Size Effects of the Exchange Energy at GTH-DZVP}
    \label{fig:dzvp-results}
\end{figure}

\subsection{Other Physical Observables}
\paragraph{Band Gaps} 
We tested the convergence of band gaps for LiH and diamond towards the TDL. While the former has a direct gap at the $X$ point, the latter has an indirect gap. Similar to the accelerated convergence demonstrated in the calculation of the exchange energy, the staggered mesh method is also expected to converge faster in computing the band gap. This is because the staggered mesh method also computes orbital energies without running into singular terms containing $|\mathbf{q}+\mathbf{G}|=\mathbf{0}$, similar to the proof of the $\mathcal{O}(N_\textbf{k}^{-1})$ (or $\mathcal{O}(N_\textbf{k}^{-1})$ under the removable discontinuity condition discussed earlier) convergence rates for exchange energy  \cite{xing_unified_2023}. We define the exchange contribution to the orbital energy of band $j$ at \textbf{k}-point $\mathbf{k}_j\in \mathcal{K}_j$ as 
\begin{align}
    \varepsilon_{K,j_{\mathbf{k}_{j}}}\left(\mathcal{K}_{i}\right)=-\frac{1}{2N_{k}}\sum_{\mathbf{k}_{i}\in\mathcal{K}_{i}}\sum_{i}\left(j_{\mathbf{k}_{j}}i_{\mathbf{k}_{i}}\mid i_{\mathbf{k}_{i}}j_{\mathbf{k}_{j}}\right)
\end{align}
where $\mathcal{K}_i$ is the \textbf{k}-mesh used to converge the SCF calculation. 
The band gap from the regular method is defined to be when the \textbf{k}-mesh for sampling the band structure is $\mathcal{K}_j=\mathcal{K}_i$. Then, we define the band gap from the staggered mesh method to be when $\mathcal{K}_i=\mathcal{K}^{1/2}_j$, that is, the \textbf{k}-mesh used to converge the SCF calculation must be a half shift from the that used to sample the band structure.
Therefore, to have a direct comparison between the regular and staggered mesh methods, the $\mathcal{K}_j$ must be the same for both. To reduce 
basis set incompleteness effects, we used the uncontracted def2-QZVP-GTH basis for band gap calculations \cite{lee_approaching_2021}.

\begin{figure}
     \centering
     \begin{subfigure}[b]{0.45\textwidth}
         \centering
         \includegraphics[width=\textwidth]{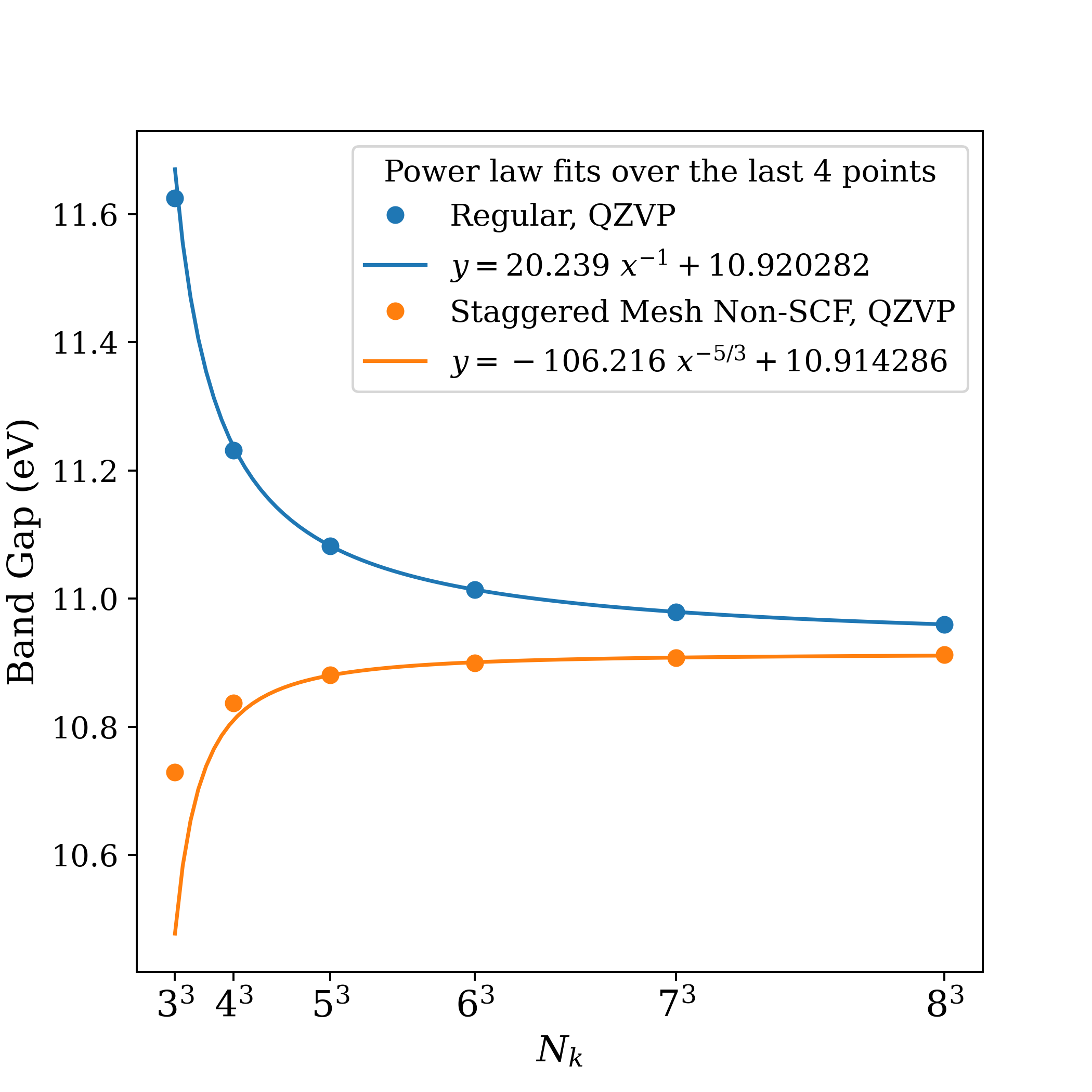}
         \caption{LiH}
         \label{fig:LiH_band_gap}
     \end{subfigure}
     \hfill
     \begin{subfigure}[b]{0.45\textwidth}
         \centering
         \includegraphics[width=\textwidth]{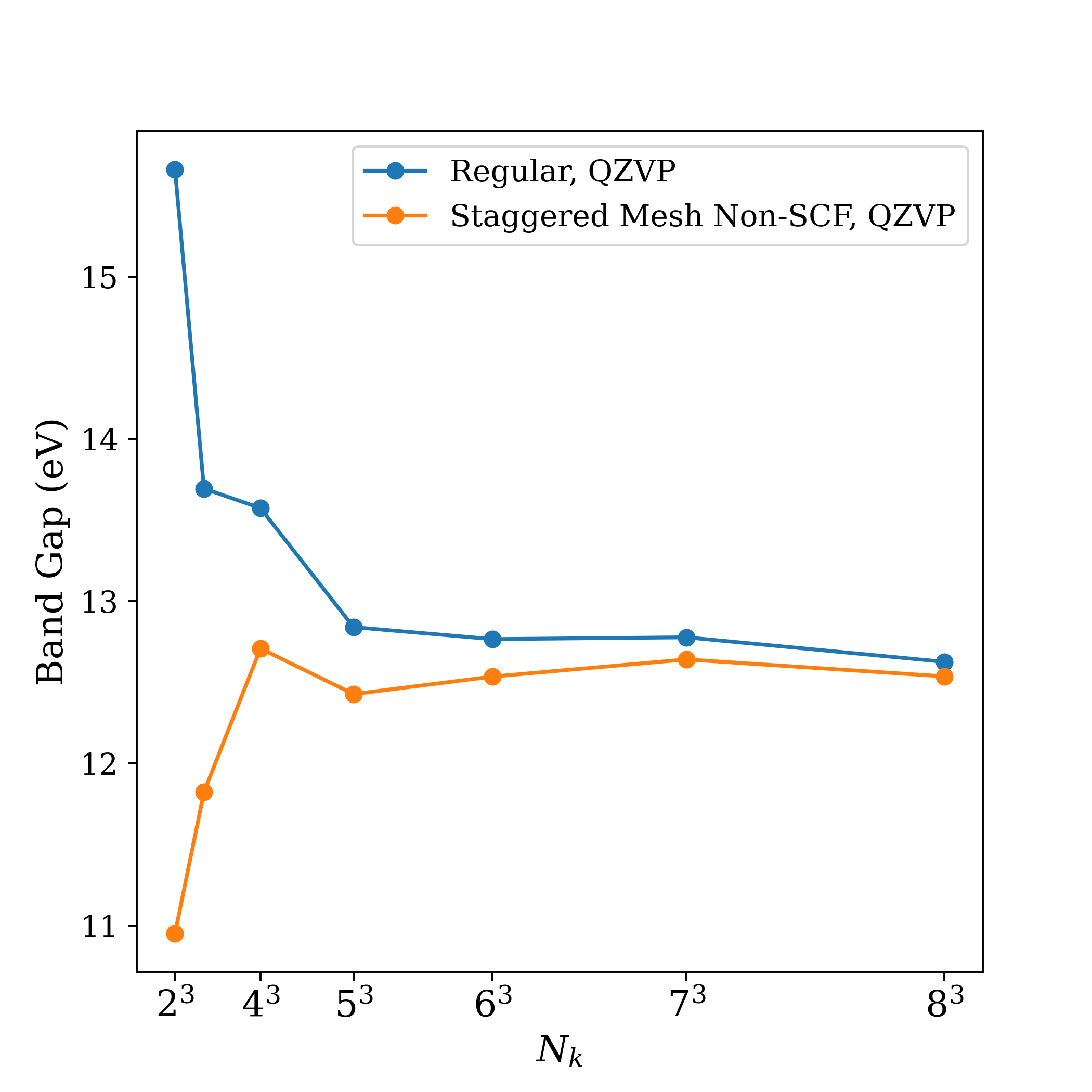}
         \caption{Diamond}
         \label{fig:diamond_band_gap}
     \end{subfigure}
        \caption{Band Gaps}
        
\end{figure}

For LiH, since the direct band gap is at the $X$ point, which corresponds to the point $\left[0.5, 0.5, 0 \right]$ in the basis of reciprocal lattice vectors,
we must set the bands k-mesh $\mathcal{K}_j$ as the $\Gamma$-centered MP mesh shifted whose points are all shifted by $\left[0.5, 0.5, 0 \right]$, and therefore, $\mathcal{K}_i$ is the $\Gamma$-centered MP mesh shifted whose points are all shifted by $\left[0, 0, 0.5 \right]$. Following the definition of the regular and staggered mesh methods above, we demonstrate the relative rates of convergence of the band gap calculation in \cref{fig:LiH_band_gap}, with good fits to polynomials of order -1 (regular) and -5/3 (staggered) respectively. The staggered mesh method is clearly superior.

The case for the staggered mesh method becomes somewhat weaker for diamond with its indirect gap. The indirect gap means it becomes more difficult to sample both the HOMO and LUMO in a single MP mesh, which can lead to oscillations as $N_\mathbf{k}$ increases as shown in \cref{fig:diamond_band_gap}. Here, we use the $\Gamma$-centered MP mesh to sample the band structure. Nevertheless, absolute errors are still lower for the staggered mesh method at nearly all $N_\mathbf{k}$.  

\paragraph{Lattice Parameters and Bulk Moduli}
We next demonstrate the effectiveness of the staggered mesh method in computing properties derived from varying the lattice parameter, $a$, of the unit cell. In particular, computing the total energy as a function of $a$ around the equilibrium lattice parameter $a_0$ and fitting a Birch-Murnaghan equation of state (EOS) \cite{birch_finite_1947,sholl_density_2009} allows one to estimate $a_0$, the equilibrium total energy $E_0$, the bulk modulus $B_0$, and the bulk modulus derivative $B'_0=(\partial B/\partial P)_T$ as parameters. The EOS is shown below. 
\begin{align}
E_{\mathrm{total}}(a)= E_0+\frac{9 V_0 B_0}{16}\left\{\left[\left(\frac{a_0}{a}\right)^2-1\right]^3 B_0^{\prime}\right. \left.+\left[\left(\frac{a_0}{a}\right)^2-1\right]^2\left[6-4\left(\frac{a_0}{a}\right)^2\right]\right\} 
\end{align}

\begin{figure}
     \centering
     \begin{subfigure}[b]{0.45\textwidth}
         \centering
         \includegraphics[width=\textwidth]{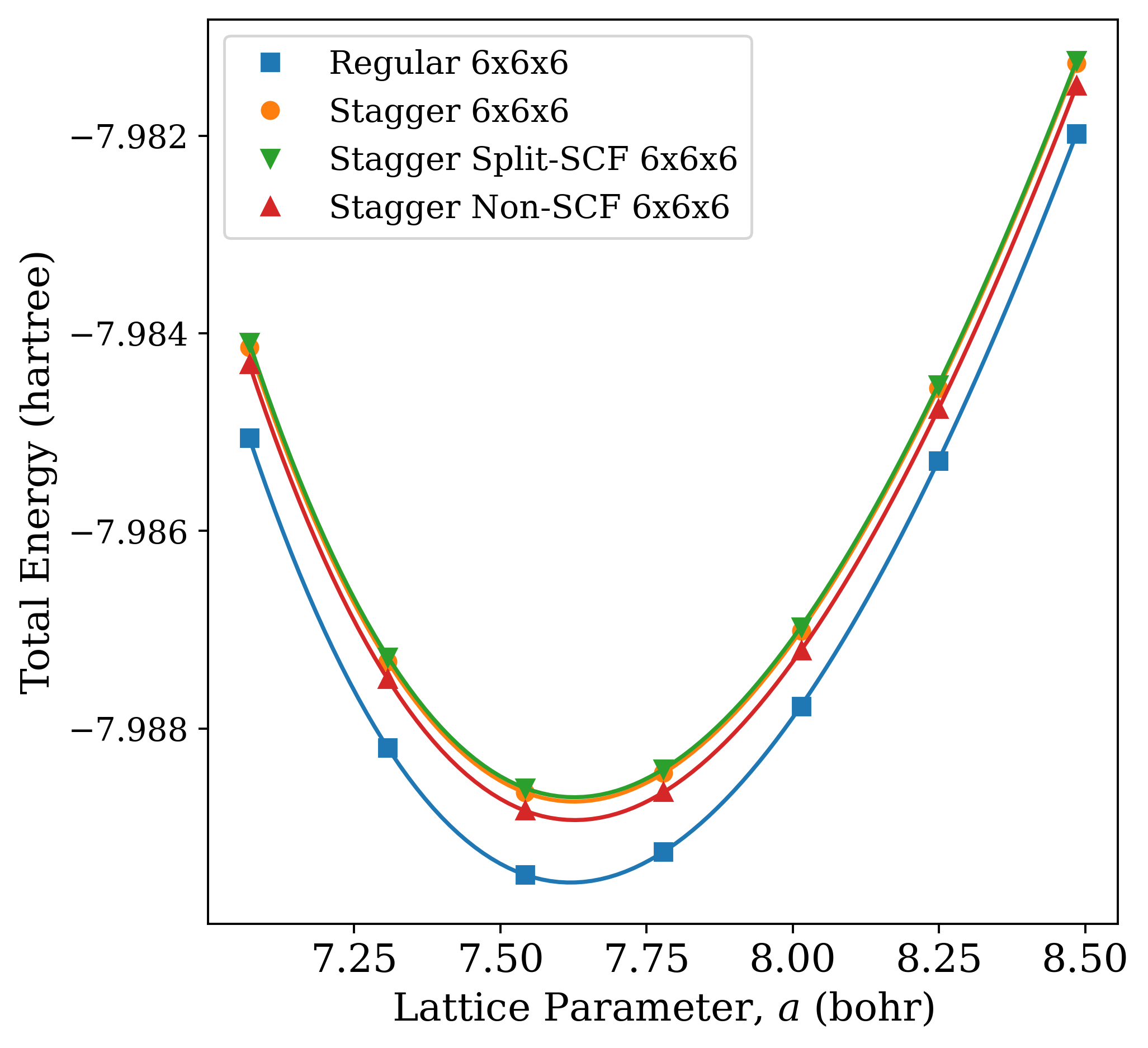}
         \caption{LiH}
         \label{fig:LiH_BM_666}
     \end{subfigure}
     \hfill
     \begin{subfigure}[b]{0.45\textwidth}
         \centering
         \includegraphics[width=\textwidth]{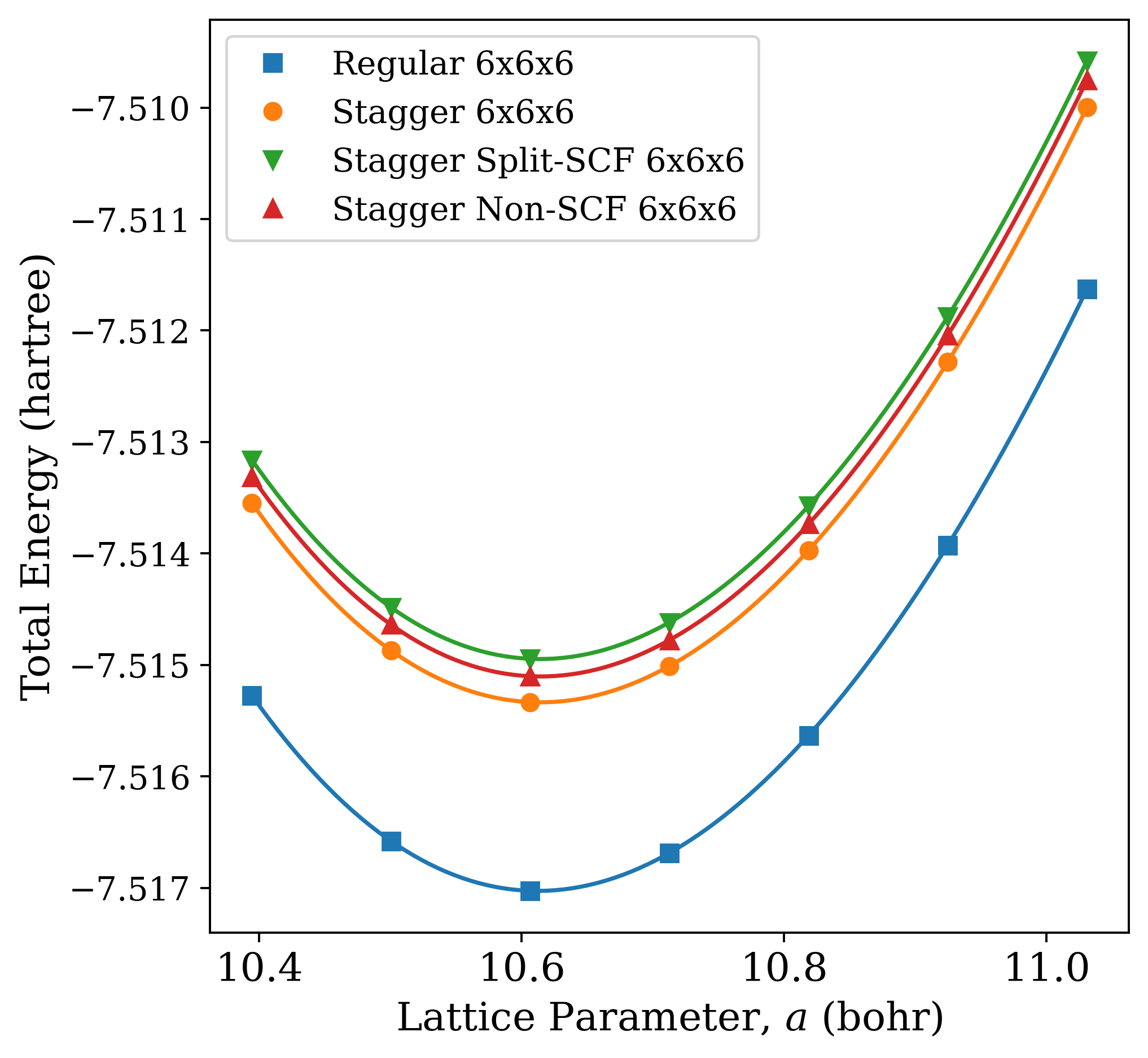}
         \caption{Si}
         \label{fig:Si_BM_666}
     \end{subfigure}
        \caption{Birch-Murnaghan Fits}
        \label{fig:all_BM_fits}
\end{figure}

\begin{figure}
     \centering
     \begin{subfigure}[b]{0.45\textwidth}
         \centering
         \includegraphics[width=\textwidth]{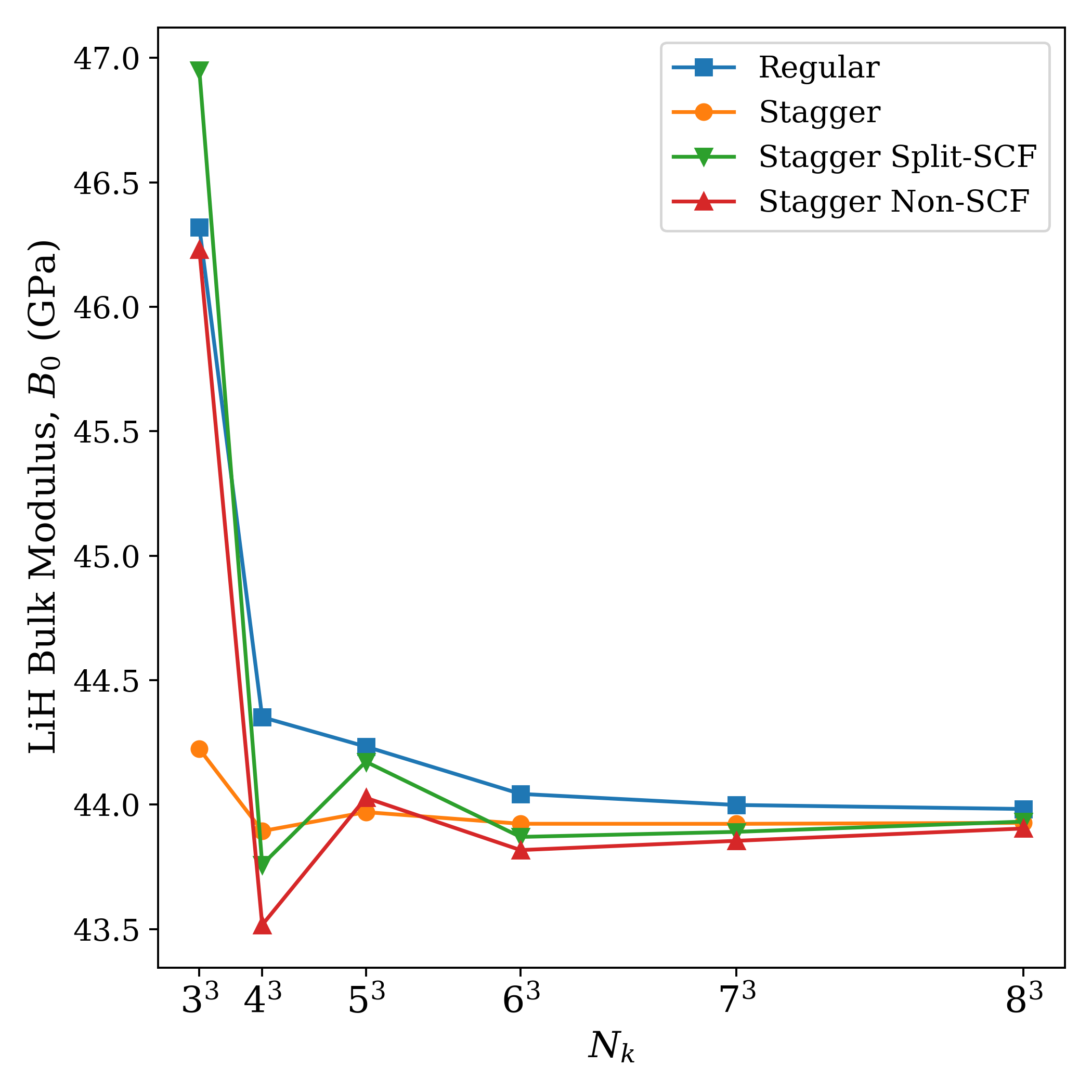}
         \caption{LiH}
         \label{fig:LiH_nk3-8_B0}
     \end{subfigure}
     \hfill
     \begin{subfigure}[b]{0.45\textwidth}
         \centering
         \includegraphics[width=\textwidth]{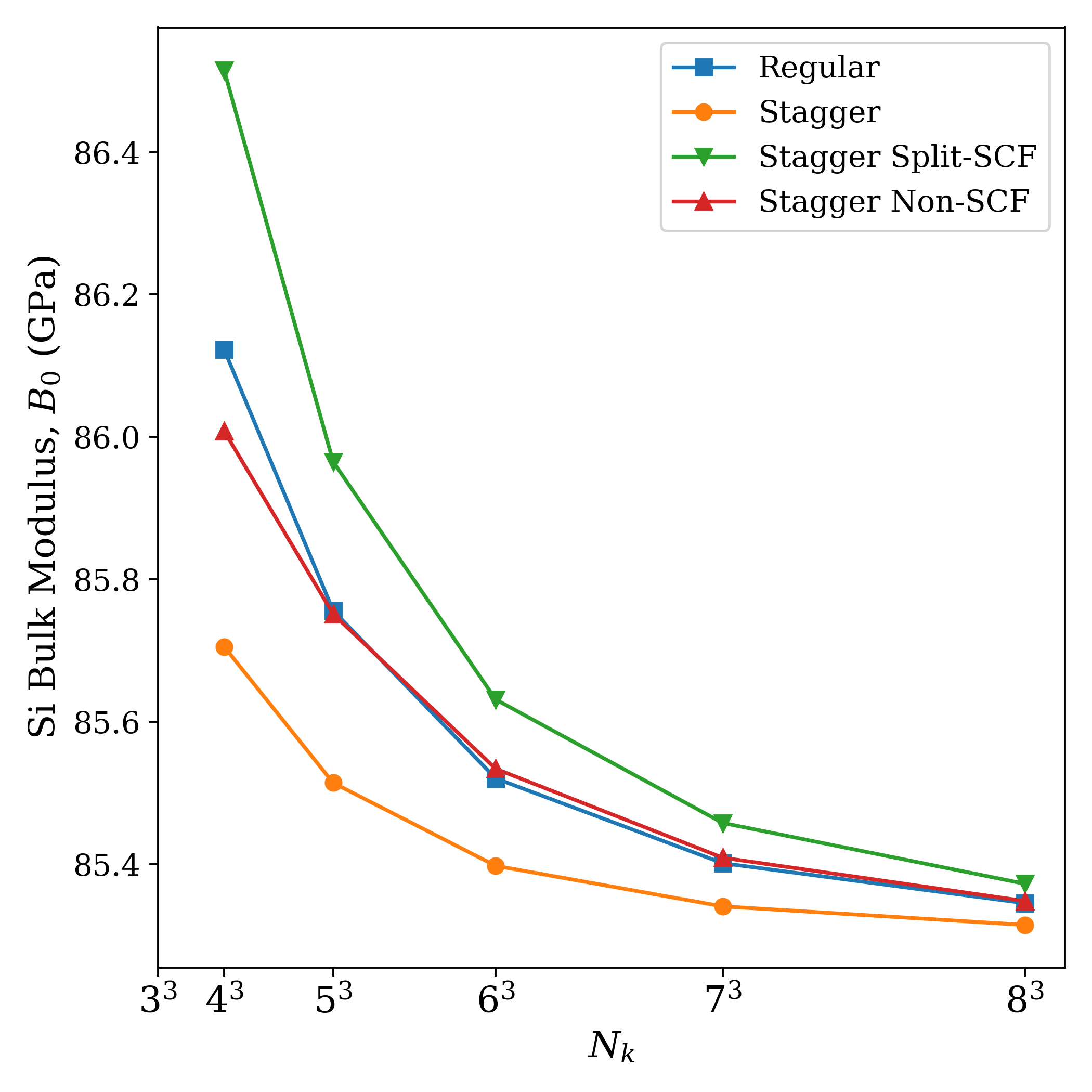}
         \caption{Si}
         \label{fig:Si_mp149_nk4-8_B0}
     \end{subfigure}
        \caption{Finite Size Effects of Bulk Modulus}
        \label{fig:all_B0_convergence}
\end{figure}

\begin{figure}
     \centering
     \begin{subfigure}[b]{0.45\textwidth}
         \centering
         \includegraphics[width=\textwidth]{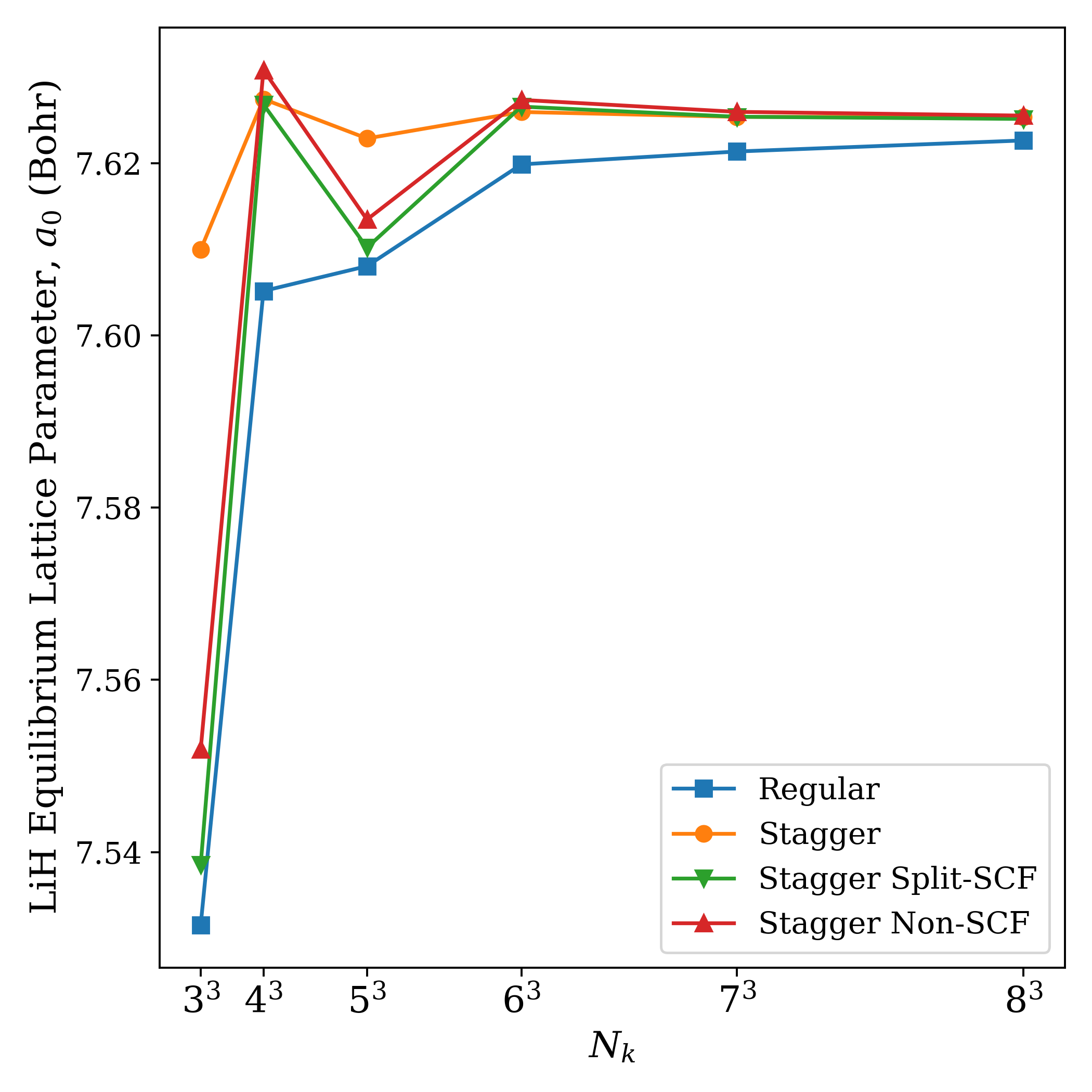}
         \caption{LiH}
         
     \end{subfigure}
     \hfill
     \begin{subfigure}[b]{0.45\textwidth}
         \centering
         \includegraphics[width=\textwidth]{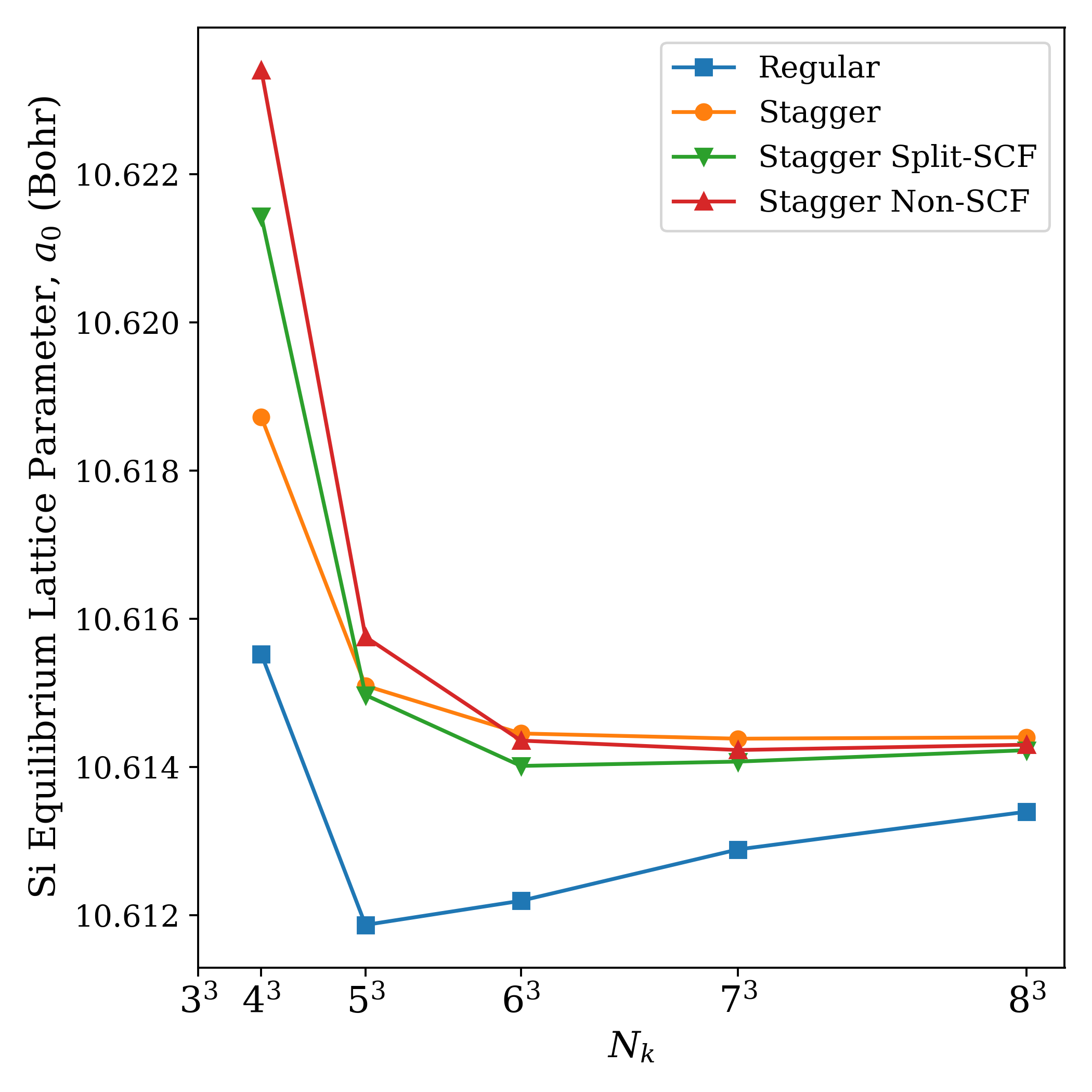}
         \caption{Si}
         \label{fig:Si_a0}
     \end{subfigure}
        \caption{Finite Size Effects of Equilibrium Lattice Parameter}
        \label{fig:all_a0_convergence}
\end{figure}

\begin{figure}
     \centering
     \begin{subfigure}[b]{0.45\textwidth}
         \centering
         \includegraphics[width=\textwidth]{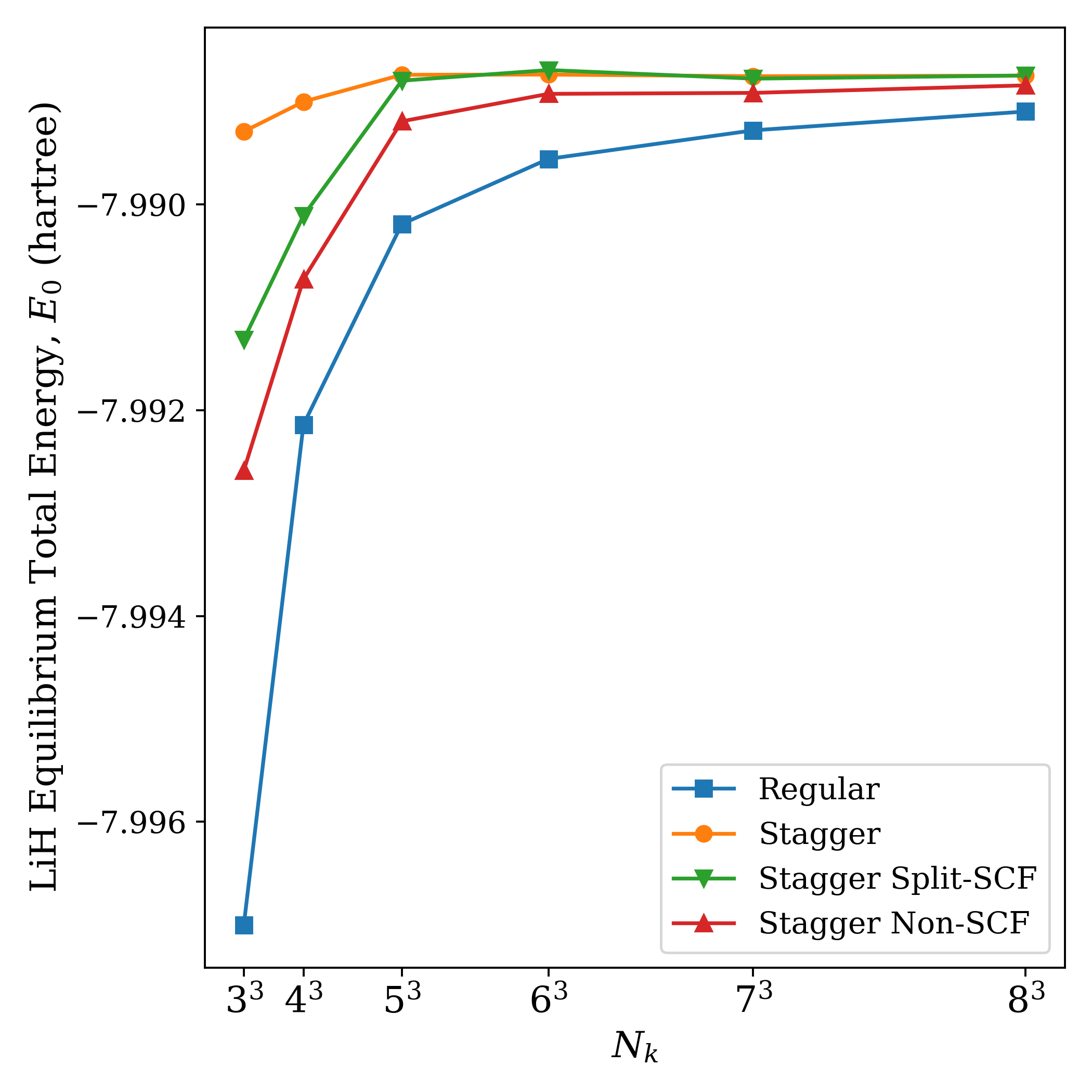}
         \caption{LiH}
         
     \end{subfigure}
     \hfill
     \begin{subfigure}[b]{0.45\textwidth}
         \centering
         \includegraphics[width=\textwidth]{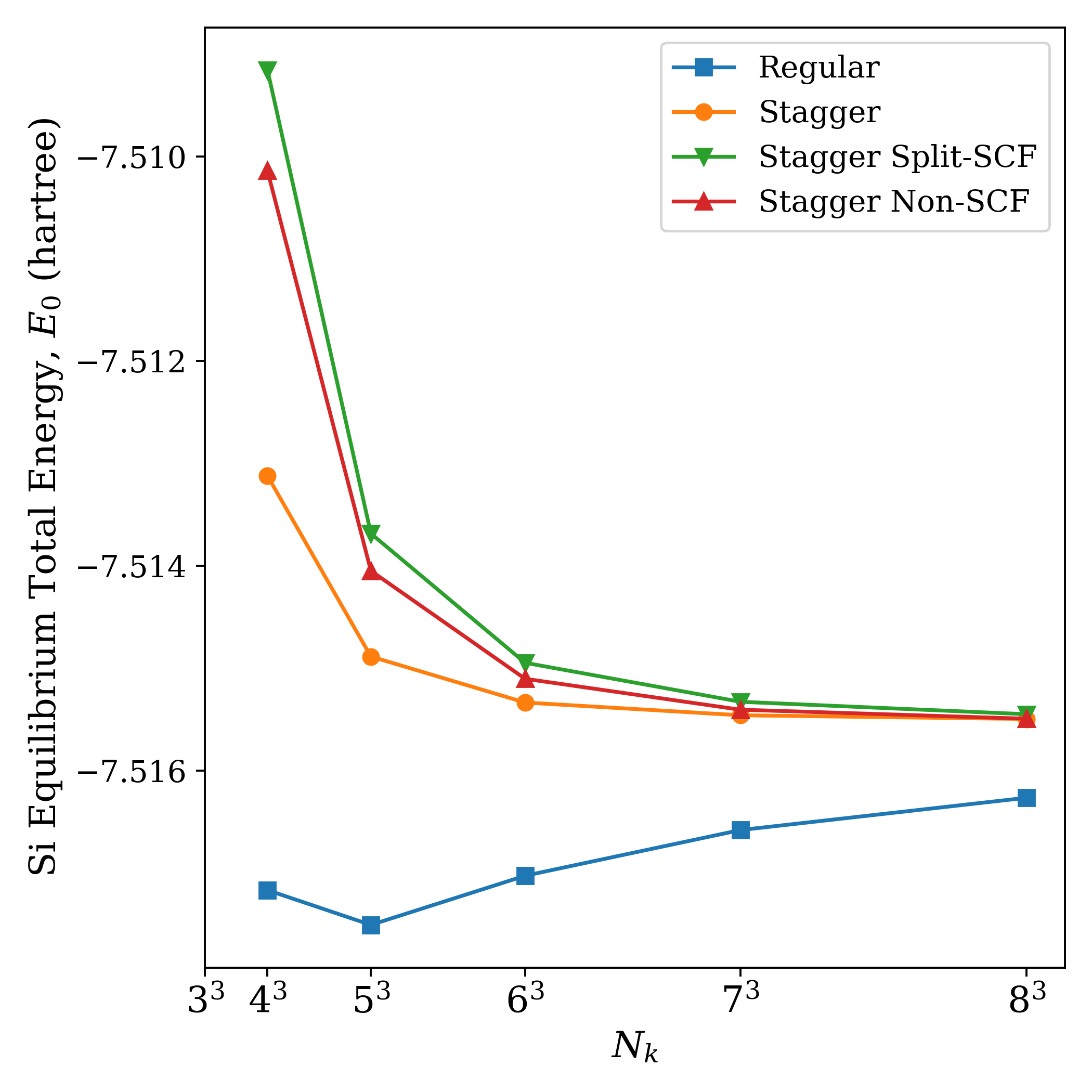}
         \caption{Si}
         \label{fig:Si_E0}
     \end{subfigure}
        \caption{Finite Size Effects of Equilibrium Total Energy}
        \label{fig:all_E0_convergence}
\end{figure}

In \cref{fig:all_BM_fits} we show the Birch-Murnaghan curves for LiH and Si using each method with $N_\mathbf{k}=6^3$. The EOS as a function of $N_\mathbf{k}$ leads to the data shown in \cref{fig:all_B0_convergence,fig:all_E0_convergence,fig:all_a0_convergence}; 
the convergence plots of $a_0$, $E_0$, and $B_0$ respectively. These plots paint a stronger picture of the difference in accuracy among the staggered mesh methods. In all plots, the original staggered mesh method performs significantly better than the Split-SCF and Non-SCF variants. Even when considering that the original variant requires double the k-points as Non-SCF, a simple comparison of the original staggered mesh at $N_\mathbf{k}=4^3=64$ versus the Split- and Non-SCF variants at $N_\mathbf{k}=5^3=125$ (or the former at  $N_\mathbf{k}=5^3$ vs. the latter at $N_\mathbf{k}=6^3=216$) shows that the original staggered mesh is still more accurate. In fact, in the convergence of $B_0$ in \cref{fig:all_B0_convergence}, the Split- and Non-SCF versions appear to do even worse than the regular method at low $N_\mathbf{k}$. 

Thus, although the Split-SCF and Non-SCF versions of staggered mesh may be more efficient in computing exchange energies, recovering measures of curvature like $B_0$ is best done through the original version of staggered mesh. We expect that the superior performance of the original version is due to (1) the invoking of a single SCF calculation on the combined mesh, which means the density matrices used in computing the exchange energy are fully self-consistent with each other, and (2) the other components of total energy such as the two-electron Coulomb matrix $J$ are also being recovered with the higher density k-mesh. 

\paragraph{Phase Transition Energies}
The ability to predict the phase transition energy between two polymorphs of a material also presents a good test for the ability of the staggered mesh method to accurately recover measurable properties that are energy differences. For this study, we look at the transition (1) from diamond to rhombohedral graphite \cite{wen_n-diamond_2006,lipson_structure_1942,white_relative_2021} and (2) from cubic to hexagonal BN \cite{cazorla_polymorphism_2019}. 
The diamond-to-graphite transition shows a particularly marked improvement for the original staggered mesh method relative to the regular method. The other two staggered mesh variants, like the regular method, display severe oscillations towards the TDL. The original staggered mesh method attenuates such oscillations, which justifies its use of double $\mathbf{k}$-points compared to a normal calculation. For the cubic-to-hexagonal BN transition, 
all three staggered mesh methods perform similarly to each other, while the regular method appears to overshoot the TDL.  Thus, just as shown in the convergence of the Burch-Murnaghan curves, all three versions of the staggered mesh method outperform the regular method, with the original version being the most suitable for these calculations. 


\begin{figure}
     \centering
     \begin{subfigure}[b]{0.45\textwidth}
         \centering
         \includegraphics[width=\textwidth]{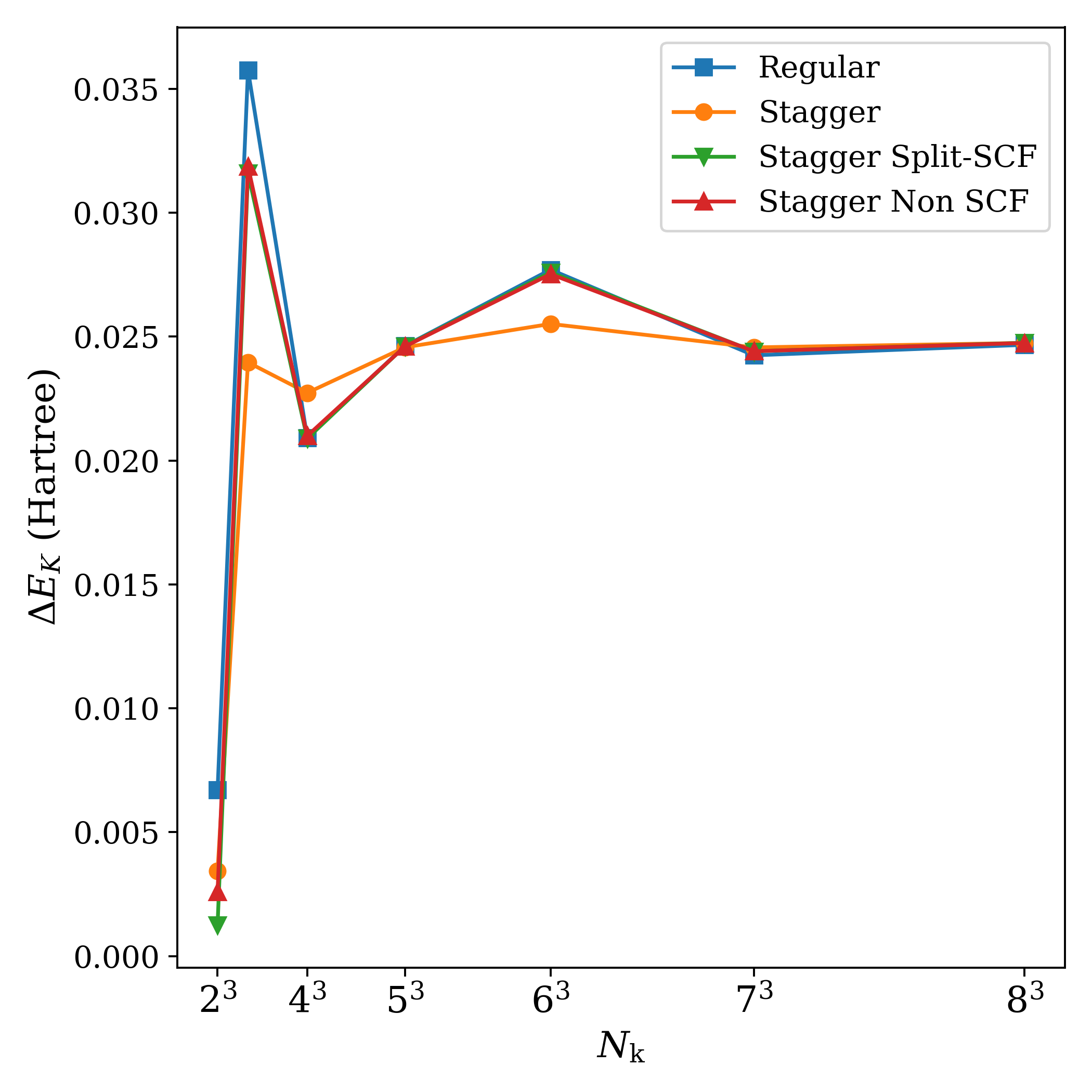}
         \caption{Diamond to Graphite}
         
     \end{subfigure}
     \hfill
     \begin{subfigure}[b]{0.45\textwidth}
         \centering
         \includegraphics[width=\textwidth]{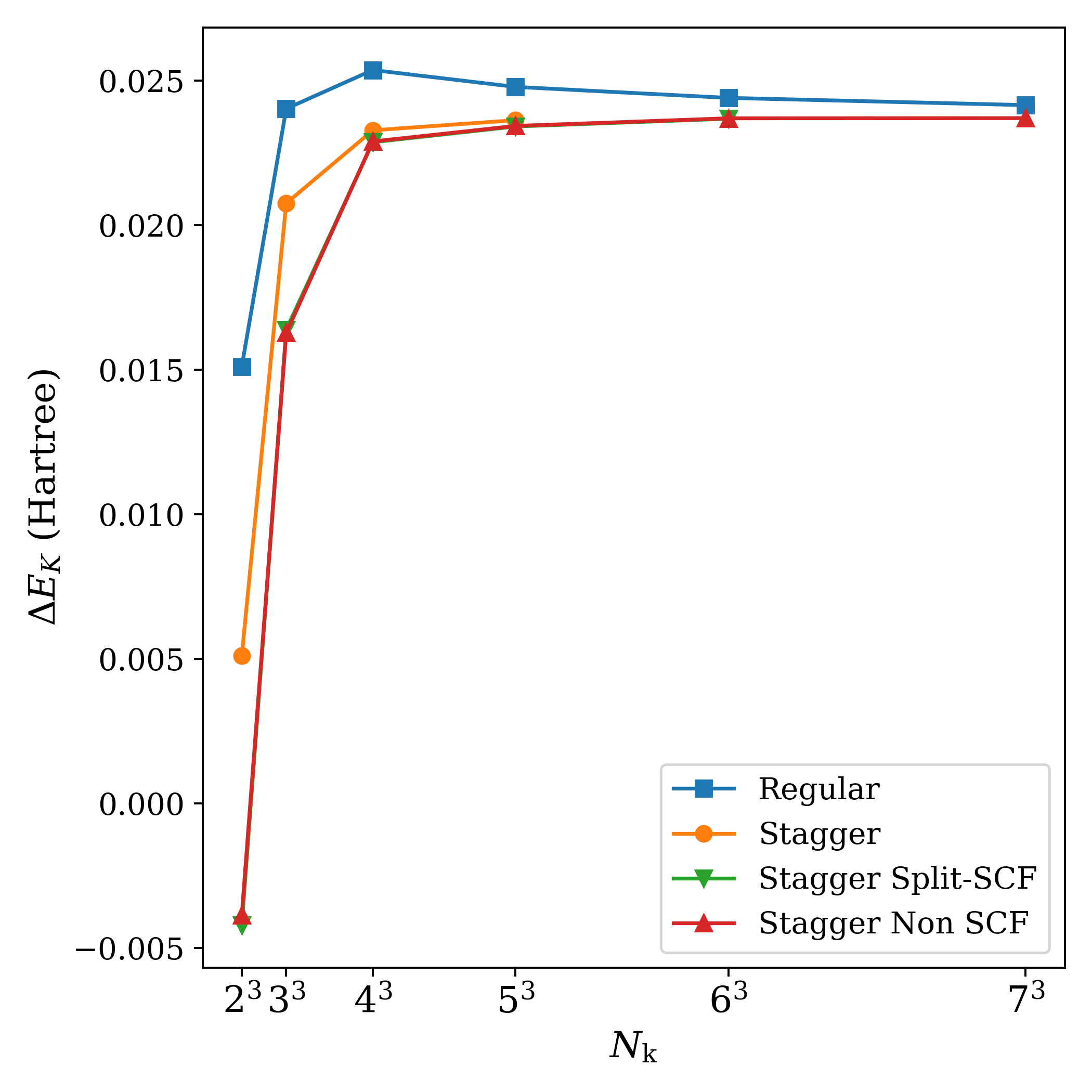}
         \caption{Cubic to Hexagonal BN. }
         \label{fig:phase_transition_bn}
     \end{subfigure}
        \caption{Finite size effects of phase transition energies. For \cref{fig:phase_transition_bn}, the incomplete data in the Stagger and Stagger Split-SCF lines is due to insufficient compute capabilities or SCF convergence issues.}
        
\end{figure}

\paragraph{Phonon Force Constants}
We test the staggered mesh method's ability to recover TDL potential energy surfaces (PES) around the equilibrium geometry for LiH and diamond. Specifically, by perturbing the atoms in the direction of a phonon, we can perform a quadratic fit of the total energy as a function of mode coordinate. For both LiH and diamond, we picked the highest frequency phonon at the $\Gamma$ point, which was verified by a Quantum Espresso calculation \cite{giannozzi_advanced_2017} and the Phonon Visualizer by Materials Cloud \cite{talirz_materials_2020}. For LiH, this corresponds to the oscillation of the H atom along the direction of a single lattice vector. We define a parametrized mode coordinate $\lambda$, so the phonon corresponds to [$\lambda$, 0.5, 0.5] in the basis of the rhombohedral primitive cell lattice vectors for LiH. Similarly, for diamond (which also has a rhombohedral primitive cell), we examine the mode corresponding to the oscillation of one of the carbon atoms along the direction of a lattice vector and define the parametrized coordinate of that atom as  [$\lambda$, 0.25 0.25]. Subsequently, $\lambda=0.5$ and $\lambda=0.25$ represent the equilibrium positions for LiH and diamond respectively.

\begin{figure}
     \centering
     \begin{subfigure}[b]{0.45\textwidth}
         \centering
         \includegraphics[width=\textwidth]{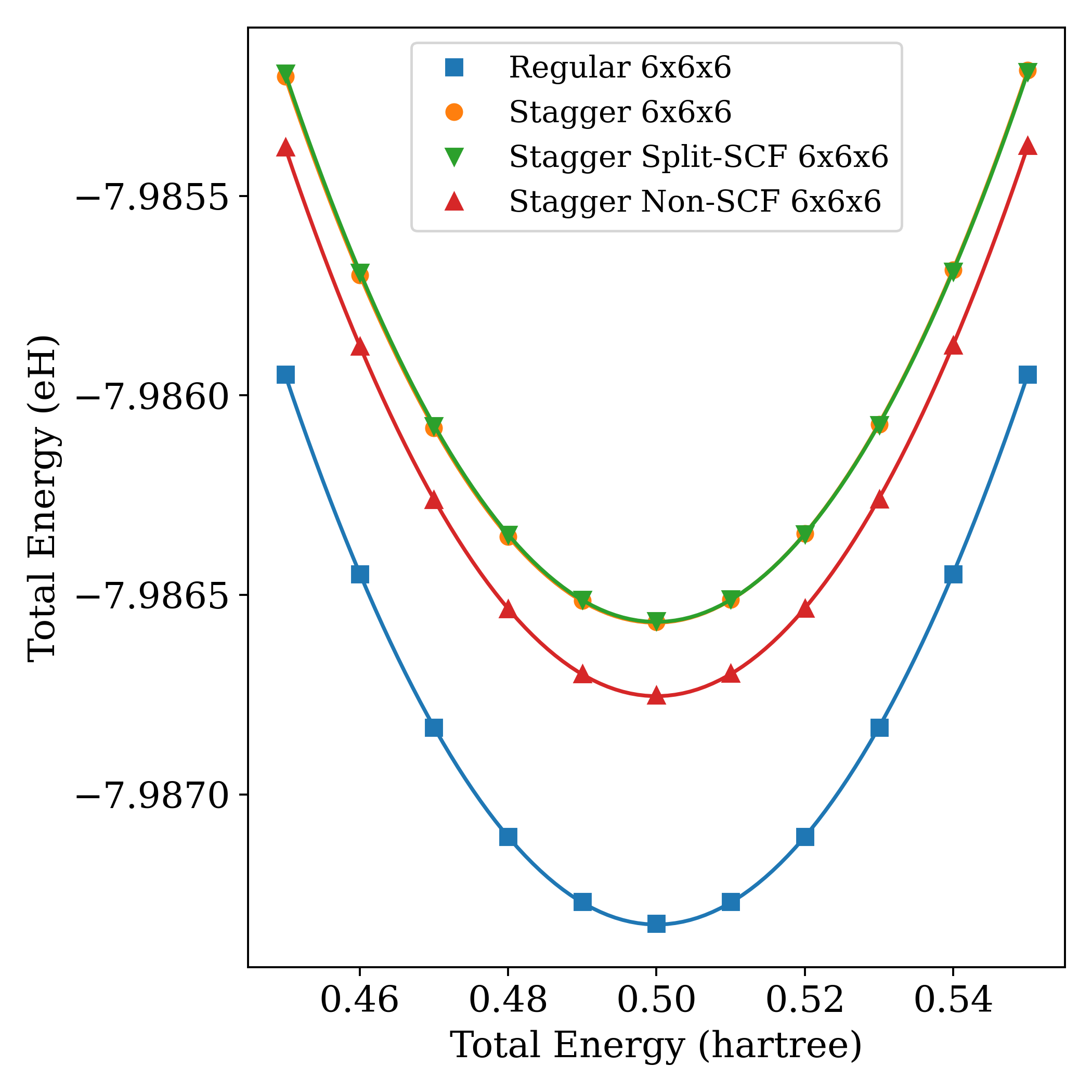}
         \caption{LiH}
         \label{fig:LiH_nk6-6_phonon-fit}
     \end{subfigure}
     \hfill
     \begin{subfigure}[b]{0.45\textwidth}
         \centering
         \includegraphics[width=\textwidth]{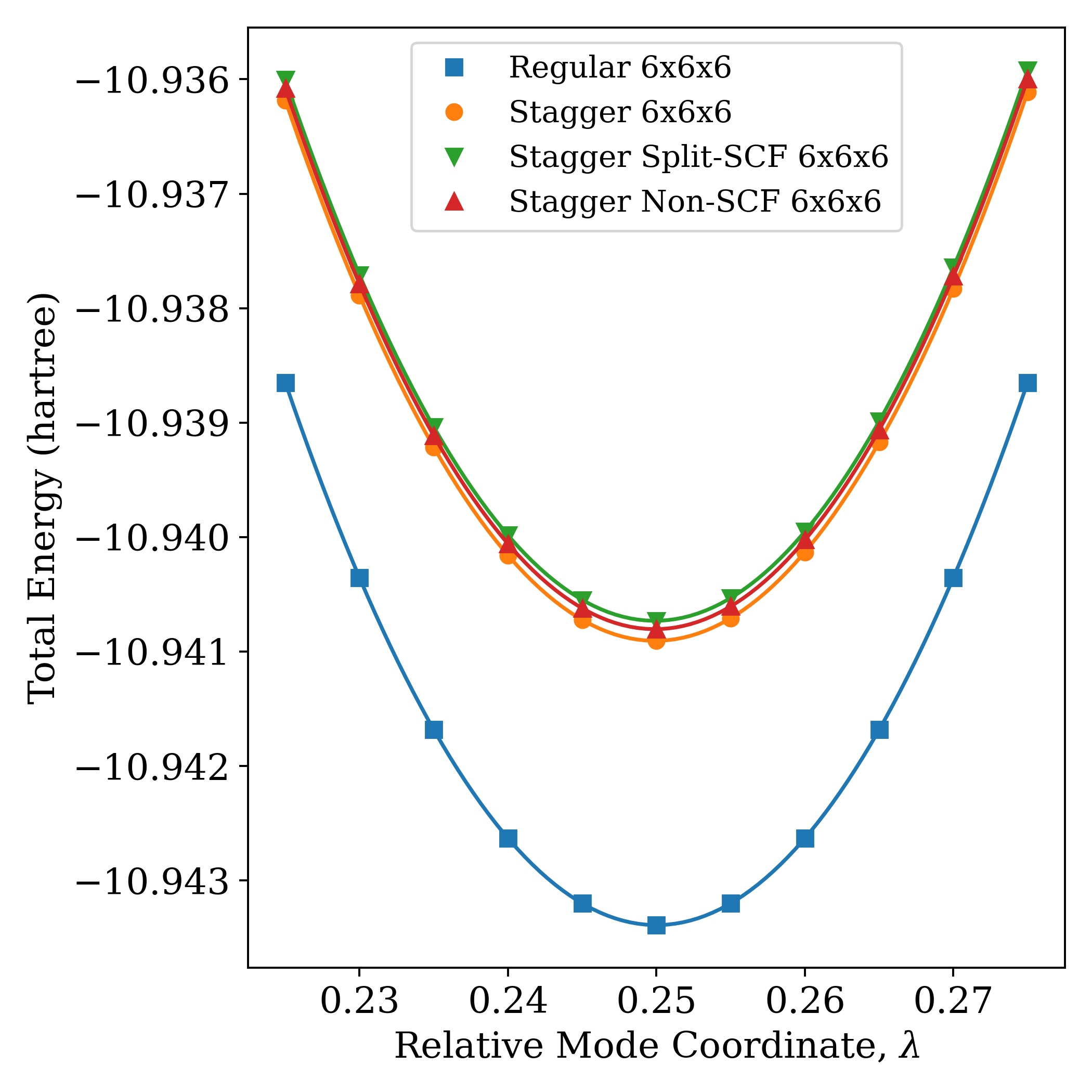}
         \caption{Diamond}
         \label{fig:diamond_mp66_nk6-6_phonon-fi}
     \end{subfigure}
        \caption{Energy Surfaces Recovered by the Regular and staggered mesh methods}
        \label{fig:all_phonon_fits}
\end{figure}

\begin{figure}
     \centering
     \begin{subfigure}[b]{0.45\textwidth}
         \centering
         \includegraphics[width=\textwidth]{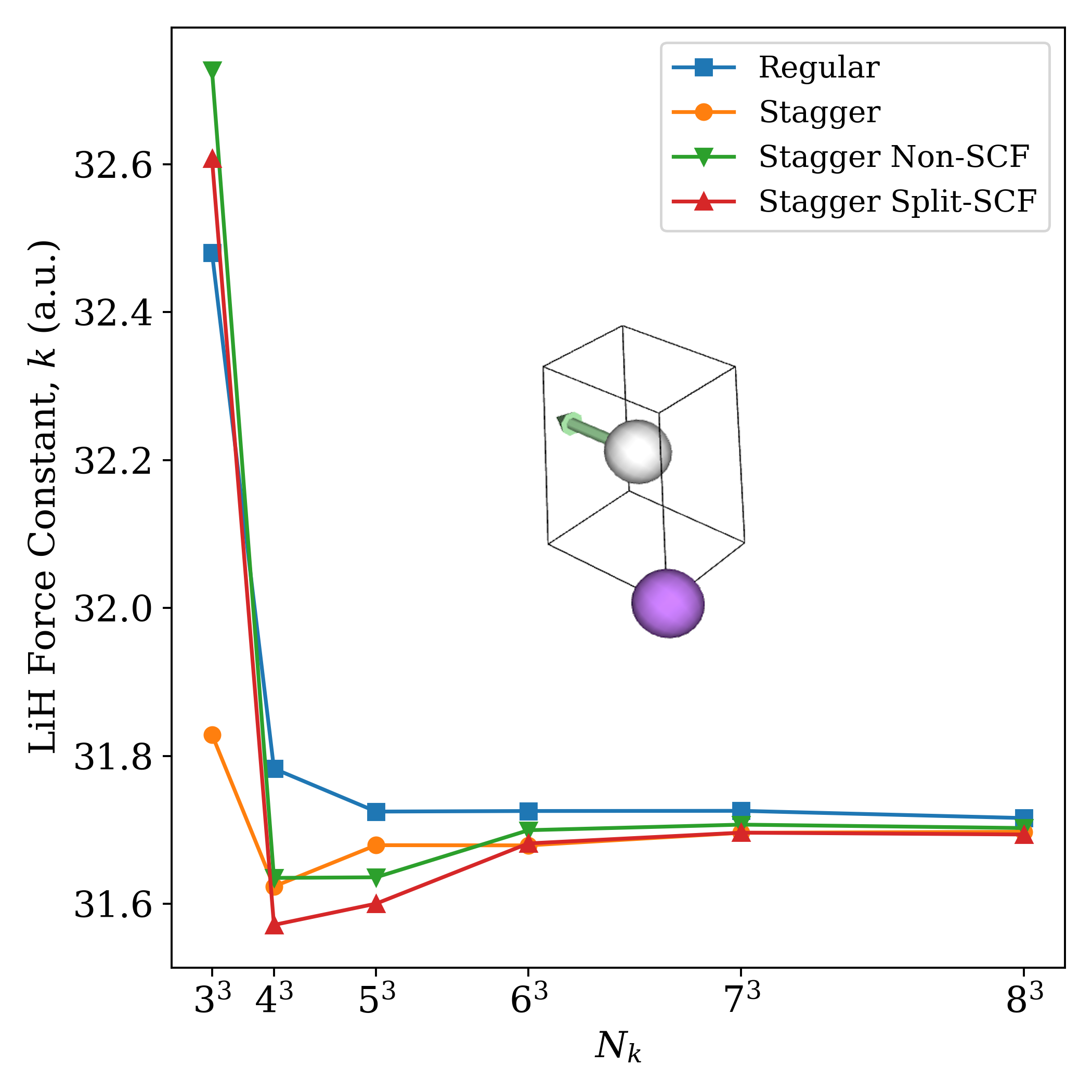}
         \caption{LiH}
         \label{fig:LiH_nk3-6_force_contant_with_mode}
     \end{subfigure}
     \hfill
     \begin{subfigure}[b]{0.45\textwidth}
         \centering
         \includegraphics[width=\textwidth]{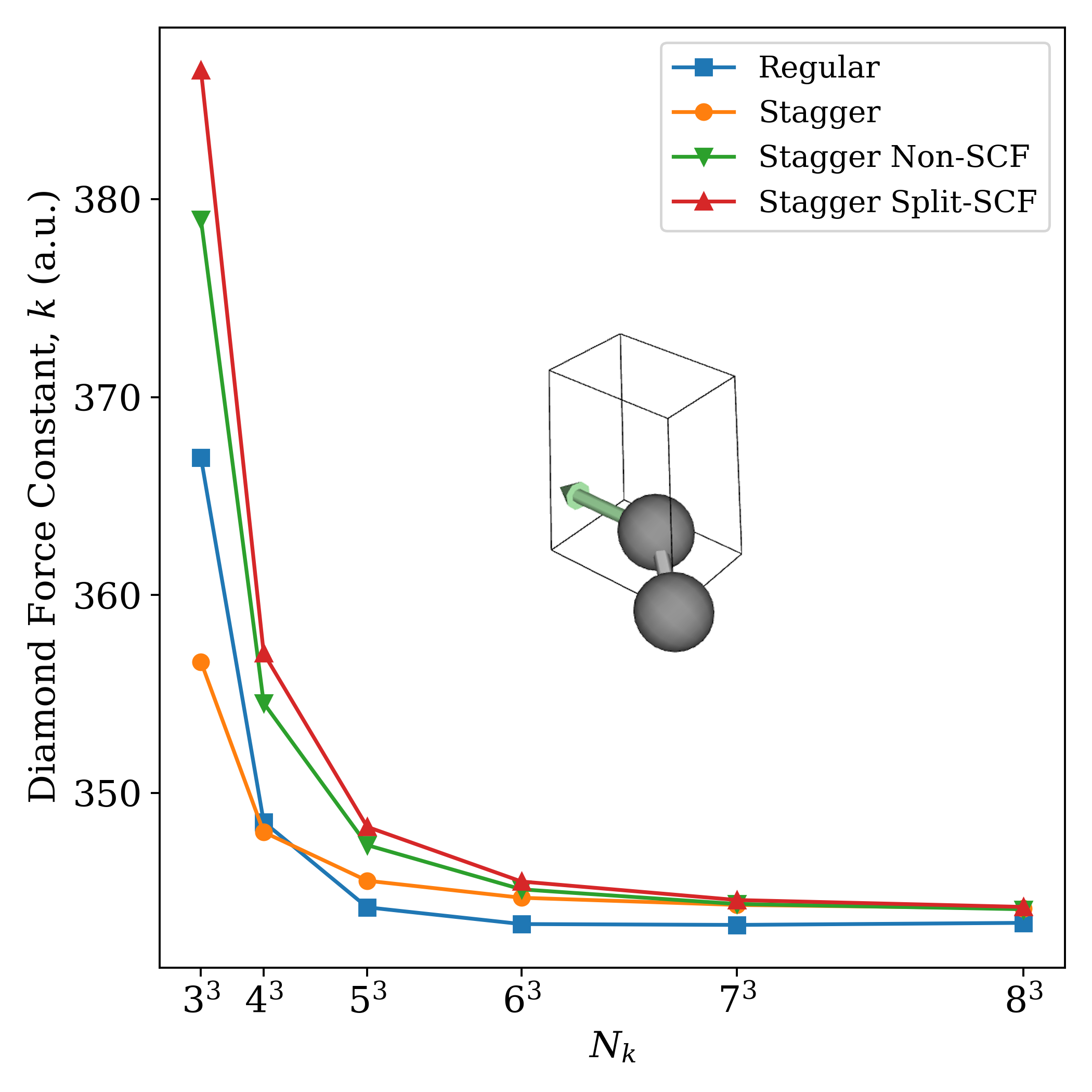}
         \caption{Diamond}
         \label{fig:diamond_mp66_nk3-6_force_constant_with_mode}
     \end{subfigure}
        \caption{Force Constant Finite Size Effect}
        \label{fig:all_force_constant_convergence}
\end{figure}

For the $6 \times 6 \times 6$ case, the PESs and their quadratic fits are shown in \cref{fig:all_phonon_fits},  while the convergence of the force constant with $N_\textbf{k}$ for each method is shown in \cref{fig:all_force_constant_convergence} with their corresponding modes visualized. The original staggered mesh method shows improvement over its two other versions as well as the regular method, especially at low $N_\textbf{k}$. We note that for diamond specifically in \cref{fig:diamond_mp66_nk3-6_force_constant_with_mode}, the force constant from the regular method may appear to be converging to a different TDL than those of the staggered mesh methods. This is because  the regular method has slightly overshot the TDL value and is still increasing even at $N_\textbf{k}=8^3$. When comparing the staggered mesh methods the Split- and Non-SCF versions only begin to perform comparably to the original staggered mesh method at high $N_\textbf{k}$. Thus, despite its slightly higher cost, the original staggered mesh method shows the most promise in computing properties that are derivatives of the total energy, such as phonon force constants. 



\section{Discussion}

The staggered mesh method presents a promising way to accelerate the convergence of the exact exchange energy towards the thermodynamic limit while requiring minimal modification to the SCF and band structure calculation procedures found in many solid-state quantum chemistry codes. It therefore has immediate applications in recovering more accurate energies obtained from using the increasingly popular hybrid functionals in periodic calculations. We also assessed the effectiveness of two other versions of the staggered mesh method designed to further reduce its cost: the Non-SCF and Split-SCF variants. We find that using staggered mesh Non-SCF recovers exchange energies and band gaps faster than the regular method as a function of $N_\textbf{k}$, at the cost of only an extra SCF cycle. In computing other properties of interest, including the bulk modulus, equilibrium lattice parameter, and equilibrium energy,  the original version performs the best.

Beyond its utility in recovering more accurate energies, the staggered mesh method points to a new technique for reducing finite-size errors in periodic calculations by avoiding unnecessary singularities. Although this work discusses only the Hartree-Fock exchange energy, the energy gradients present a natural next step for testing the effectiveness of the staggered mesh method for a range of realistic systems. Staggered mesh techniques have also been developed for Second Order Moller-Plesset Perturbation theory (MP2) \cite{xing_staggered_2021}, coupled cluster, and random phase approximation (RPA) calculations\cite{xing_staggered_2022}. We also aim to extend the staggered mesh method to MP3 and its improved variants, in conjunction with acceleration techniques, such as tensor hypercontraction (THC) \cite{bertels_third-order_2019}. While the present work primarily deals with 3D extended systems, studies on 2D (e.g. graphene, hexagonal BN) and 1D extended systems (polymers, nanotubes) \cite{shanmugam_review_2022,faulstich_interacting_2023,maschio_direct_2018,ruiperez_application_2019,kastinen_intrinsic_2016} also present an opportunity for rigorous testing of the staggered mesh method in the wavefunction methods mentioned above.
\begin{acknowledgement}

This material is based upon work supported by the National Science Foundation Graduate Research Fellowship Program under Grant No. DGE 2146752 (SJQ) and the U.S. Department of Energy, Office of Science, Office of Advanced Scientific Computing Research and Office of Basic Energy Sciences, Scientific Discovery through Advanced Computing (SciDAC) program under Award Number DE-SC0022364 (LL, MHG). LL is a Simons Investigator in Mathematics.


\end{acknowledgement}

\begin{suppinfo}

Raw data for exchange energies, total energies, band gaps, Birch-Murnaghan fit data, and phonon data for the systems discussed in the work. 

\end{suppinfo}

\bibliography{Staggered-Mesh-0224}

\end{document}